\documentclass[letterpaper,twocolumn,10pt]{article}
\usepackage{usenix}
\usepackage[available,functional,reproduced]{usenixbadges}

\usepackage{tikz}
\usepackage{float}
\usepackage{amsmath}
\usepackage{balance}
\usepackage{booktabs}
\usepackage[capitalise]{cleveref}
\usepackage{xcolor}
\usepackage{xspace}
\usepackage{multirow}
\usepackage{tabularx}
\usepackage{algorithm}
\usepackage{algpseudocode}
\algtext*{EndWhile}
\algtext*{EndIf}
\algtext*{EndFor}
\usepackage{caption}
\usepackage{subcaption}
\usepackage[compact]{titlesec}

\usepackage{enumitem}
\setitemize{noitemsep,topsep=0pt,parsep=0pt,partopsep=0pt}

\newif\ifcomments
\commentstrue 
\ifcomments
    \newcommand{\revision}[1]{{\color{blue}{\bf\sf #1}}}
\else
    \newcommand{\revision}[1]{ #1}
\fi

\newcommand{\sys}[0]{DistServe\xspace}
\def\Snospace~{\S{}}

\newcommand{\parabf}[1]{\medskip\noindent\textbf{#1}}

\newcommand{\paraf}[1]{\noindent\textbf{#1}}
\newcommand{\cut}[1]{}

\begin{document}

\date{}

\title{\sys: Disaggregating Prefill and Decoding for Goodput-optimized \\
Large Language Model Serving}

\author{
\rm{Yinmin Zhong$^{\text{1}}$ \enskip
    Shengyu Liu$^{\text{1}}$ \enskip
    Junda Chen$^{\text{3}}$ \enskip
    Jianbo Hu$^{\text{1}}$ \enskip
    Yibo Zhu$^{\text{2}}$  \enskip
    Xuanzhe Liu$^{\text{1}}$ \enskip
    }
\\
\rm{
    Xin Jin$^{\text{1}}$ \enskip
    Hao Zhang$^{\text{3}}$ \enskip
    }\\
\\
{$^{\text{1}}$School of Computer Science, Peking University\enskip $^{\text{2}}$StepFun\enskip $^{\text{3}}$UC San Diego\enskip}
}

\maketitle

\begin{abstract}
\sys improves the performance of large language models (LLMs) serving by disaggregating the prefill and decoding computation. Existing LLM serving systems colocate the two phases and batch the computation of prefill and decoding across all users and requests. 
We find that this strategy not only leads to strong prefill-decoding interferences but also couples the resource allocation and parallelism plans for both phases. LLM applications often emphasize individual latency for each phase: time to first token (TTFT) for the prefill phase and time per output token (TPOT) of each request for the decoding phase.
In the presence of stringent latency requirements, existing systems have to prioritize one latency over the other, or over-provision compute resources to meet both. 

\sys assigns prefill and decoding computation to different GPUs, hence eliminating prefill-decoding interferences. Given the application's TTFT and TPOT requirements, \sys co-optimizes the resource allocation and parallelism strategy \emph{tailored} for each phase. \sys also places the two phases according to the serving cluster's bandwidth to minimize the communication caused by disaggregation. As a result, \sys significantly improves LLM serving performance in terms of the maximum rate that can be served within both TTFT and TPOT constraints on each GPU.
Our evaluations show that on various popular LLMs, applications, and latency requirements, \sys can serve 7.4$\times$ more requests or 12.6$\times$ tighter SLO, compared to state-of-the-art systems, while staying within latency constraints for $>90\%$ of requests.
    
\end{abstract}
\section{Introduction}
\label{sec:introduction}

Large language models (LLMs), such as GPT-4~\cite{openai2023gpt4}, Bard~\cite{bard}, and LLaMA~\cite{touvron2023llama}, represent a groundbreaking shift in generative AI. They start to reshape existing Internet services, ranging from search engines to personal assistants~\cite{inflection}, and enable fundamentally new applications, like universal chatbots~\cite{chatgpt,vicuna2023} and programming assistants~\cite{chen2021evaluating,roziere2023code}. Yet, these advances come with a significant challenge: processing an end-to-end LLM query can be substantially slower than a standard search query~\cite{chat-cost}. In order to meet the stringent latency requirements of various applications, service providers need to over-provision compute resources, particularly many GPUs, leading to a shortfall in cost efficiency. Therefore, optimizing the cost per LLM query while adhering to high \textit{SLO attainment} (the proportion of requests that meet the SLOs) is becoming increasingly essential for all LLM services.

\begin{figure}[t]
    \centering
    \includegraphics[width=0.8\columnwidth]{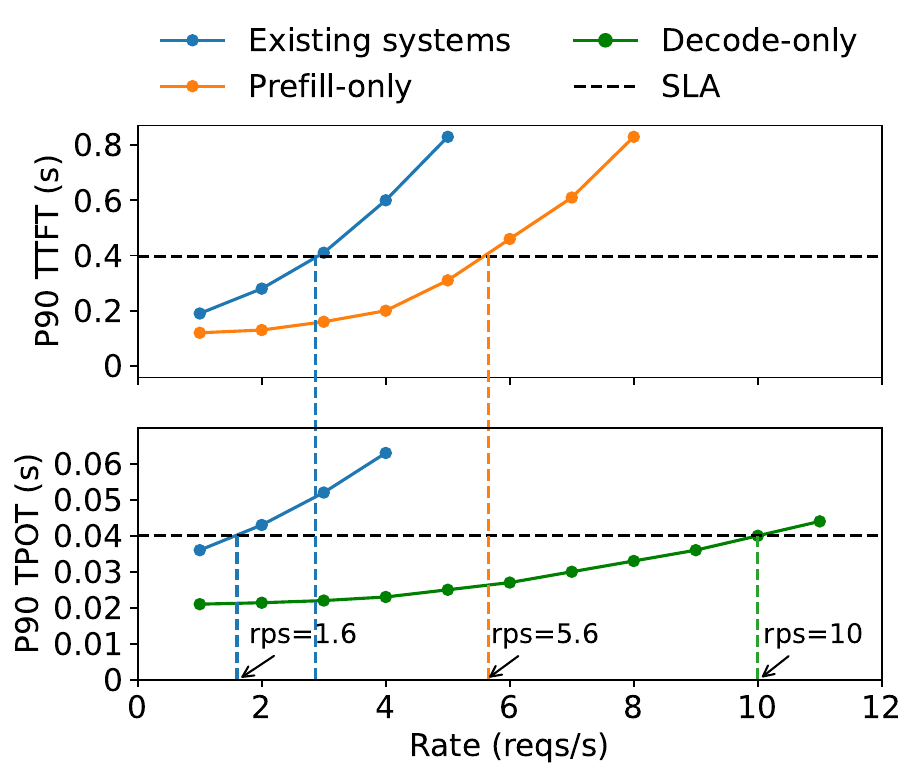}
    \vspace{-10pt}
    \caption{
    Performance when serving an LLM with 13B parameters under a synthetic workload with input length = 512 and output length = 64 on one NVIDIA 80GB A100. \textit{Upper}: The P90 time-to-first-token (TTFT) latency comparing existing systems vs. a system serving only the prefill phase. \textit{Down}: The P90 time-per-output-token (TPOT) latency comparing existing systems vs. a system serving only the decoding phase.}
    \label{fig:teaser}
    \vspace{-15pt}
\end{figure}

An LLM service responds to a user query in two phases. The \emph{prefill phase} processes a user's prompt, composed of a sequence of tokens, to generate the first token of the response \emph{in one step}. Following it, the \emph{decoding phase} sequentially generates subsequent tokens \emph{in multiple steps}; each decoding step generates a new token based on tokens generated in previous steps, until reaching a termination token.
This dual-phase process distinguishes LLM services from traditional services
-- an LLM service's latency is uniquely measured by two key metrics: the \emph{time to first token} (TTFT), which is the duration of the prefill phase, and the \emph{time per output token} (TPOT), which represents the average time taken to generate a token for each request (except for the first token)\footnote{The overall request latency equals TTFT plus TPOT times the number of generated tokens in the decoding phase.}. 
Different applications place varying demands on each metric. For example, real-time chatbots~\cite{chatgpt} prioritize low TTFT for response promptness, while TPOT only remains important until it is faster than human reading speed (i.e., 250 words/min).
Conversely, document summarization emphasizes low TPOT for faster generation of the summary.

Hence, given the application's TTFT and TPOT requirements, an effective LLM serving system should balance these needs and maximize \emph{per-GPU goodput}, defined as the maximum request rate that can be served adhering to the SLO attainment goal (say, 90\%) for each GPU provisioned -- higher per-GPU goodput directly translates into lower cost per query.

As the prefill and decoding phases share the LLM weights and working memory,
existing LLM serving systems typically colocate both phases on GPUs and maximize the overall system throughput  -- tokens generated per second across all users and requests -- by batching the prefill and decoding steps across requests~\cite{yu2022orca,kwon2023efficient}. However, to meet latency requirements, we find these systems must over-provision compute resources. To see this, Figure~\ref{fig:teaser} illustrates how the P90 TTFT and TPOT shift with increasing request rates when serving a 13B LLM using existing systems~\cite{vllm}, with workload pattern and two latency constraints set to emulate using LLM to generate a short summary for an article. Under the SLO attainment of 90\%, the maximum achievable goodput on a single A100 GPU, which is constrained by the more stringent one of TTFT and TPOT requirements, is about 1.6 requests per second (rps). 
The performance contrasts sharply when each phase is served independently on a separate GPU, shown by the orange and green curves, which achieve per-GPU goodput of 5.6 rps for the prefill phase and 10 rps for decoding. Ideally, by allocating 2 GPUs for prefill and 1 GPU for decoding, we can effectively serve the model with an overall goodput of 10 rps, or equally 3.3 rps per GPU, which is 2.1x higher than existing systems.
The gap in goodput primarily stems from the colocation of the prefill and decoding  -- two phases with very distinct computational characteristics and latency requirements (\S\ref{sec:llm-inference}).

First, colocation leads to strong \emph{prefill-decoding interference}.
A prefill step often takes much longer than a decoding step. When batched together, decoding steps in the batch are delayed by the prefill steps, significantly elongating their TPOT; similarly, the inclusion of decoding steps contributes to a non-trivial increase in TTFT, as evidenced in Figure~\ref{fig:interference}.
Even if we schedule them separately, issues persist as they begin to compete for resources. Decoding tasks awaiting GPU execution are subject to increased queuing delays due to ongoing prefill tasks, and vice versa. Prioritized scheduling of one phase risks failing the latency requirements of the other.

Second, the prefill and decoding computation differ in latency requirements and preference for different forms of parallelism (\S\ref{sec:tradeoff}). Colocating prefill and decoding, however, couples their resource allocation, and prevents implementing different parallelism strategies more suited to meeting the specific latency requirements of each phase. 

To overcome these challenges, we propose to disaggregate the prefill and decoding phases of LLM inference, assigning them to separate GPUs. Our approach has two benefits.
First, operating each phase independently on different GPUs eliminates prefill-decoding interference. Second, it allows to scale each phase independently with tailored resource allocation and model parallelism strategies to meet their specific latency requirements.
Although disaggregation causes communication of intermediate states between GPUs, we show that the communication overhead is insubstantial (\S\ref{subsec:practical_problems}) in modern GPU clusters, and when managed appropriately, disaggregation significantly improves per-GPU goodput. 

Based on the above insights, in this work, we build \sys\footnote{\revision{\href{https://github.com/LLMServe/DistServe}{https://github.com/LLMServe/DistServe}}}, a goodput-optimized LLM serving system by disaggregating the prefill and decoding phases. Given TTFT and TPOT requirements, \sys first scales each phase independently by co-optimizing the GPU allocation and parallelism strategies of the prefill and decoding phase assuming serving a single model replica. The optimization ensures maximizing the per-GPU goodput and may assign different numbers of GPUs and parallelism strategies to each phase depending on their respective latency requirements. \sys then scales this allocation to multiple instances via replication until meeting the user-required traffic rate (\S\ref{sec:method}). 
\sys also features an algorithm to place the prefill and decoding computation according to their allocation schemes and the cluster's bandwidth to minimize the overhead of communicating intermediate states between phases.

We implement \sys as an orchestration layer on top of the LLM inference engine. We
evaluate \sys on various LLMs, varying the workloads based on three important real-world LLM applications: chatbots, programming assistant, and document summary. Compared to state-of-the-art solutions, \sys can serve up to $7.4\times$ more requests or $12.6\times$ tighter SLO under various latency constraints. Our contributions are:

\begin{itemize}
    \item Identify the problems of prefill-decoding interference and resource coupling in existing LLM serving systems and propose to disaggregate the two phases.
    \item Design a novel placement algorithm to choose the goodput-optimal schema for prefill and decoding instances automatically.
    \item Conduct a comprehensive evaluation of \sys with realistic workloads.
\end{itemize}

\section{Background and Motivation}
\label{sec:background}

An LLM service follows a client-server architecture: the client submits a sequence of text as a request to the server; the server hosts the LLM on GPUs, runs inference over the request, and responds (or streams) the generation back to the client. As explained in \S\ref{sec:introduction}, due to the unique prefill-decoding process, LLM service may impose aggressive service-level objectives (SLOs) on both TTFT and TPOT, varying with the application's needs. The serving system must meet both SLOs while minimizing the cost associated with expensive GPUs. In other words, we want the serving system to maximize the requests served per second adhering to the SLO attainment goal for each GPU provisioned -- \emph{maximizing per-GPU goodput}.
Next, we detail the LLM inference computation (\S\ref{sec:llm-inference}) and discuss existing optimizations for LLM serving (\S\ref{sec:llm-serving}).

\subsection{LLM Inference}
\label{sec:llm-inference}

Modern LLMs~\cite{openai2023gpt4,touvron2023llama} predict the next token given an input sequence. This prediction involves computing a hidden representation for each token within the sequence. An LLM can take a variable number of input tokens and compute their hidden representations in parallel, and its computation workload increases superlinearly with the number of tokens processed in parallel. Regardless of the input token count, the computation demands substantial I/O to move LLM weights and intermediate states from the GPU's HBM to SRAM. This process is consistent across varying input sizes.

The prefill step deals with a new sequence, often comprising many tokens, and processes these tokens concurrently. Unlike prefill, each decoding step only processes one new token generated by the previous step.
This leads to significant computational differences between the two phases.
When dealing with user prompts that are not brief, the prefill step tends to be compute-bound. For instance, for a 13B LLM, computing the prefill of a 512-token sequence makes an A100 near compute-bound (see \S\ref{subsec:analysis_prefill}). In contrast, despite processing only one new token per step, the decoding phase incurs a similar level of I/O to the prefill phase, making it constrained by the GPU's memory bandwidth.

During both phases, intermediate states, known as KV caches~\cite{vllm}, are generated at each token position, which are needed again in later decoding steps. To avoid recomputing them, they are saved in GPU memory. 
Because of the shared use of LLM weights and KV caches in memory, most LLM inference engines opt to colocate the prefill and decoding phases on GPUs, despite their distinct computational characteristics.

\subsection{LLM Serving Optimization}
\label{sec:llm-serving}
In real-time online serving, multiple requests come and must be served within SLOs. Batching and parallelizing their computation is key for achieving low latency, high throughput, and high utilization of GPUs. 

\parabf{Batching.} 
Current serving systems~\cite{yu2022orca,vllm, agrawal2023sarathi} utilize a batching technique known as \emph{continuous batching}. This method batches the prefill of new requests with the decoding of ongoing ones. It boosts the GPU utilization and maximizes the overall system throughput -- tokens generated per second across all users and requests.
However, as mentioned in \S\ref{sec:introduction} and elaborated later in \S\ref{sec:problem-opportunity}, this approach leads to trade-offs between TTFT and TPOT.
An advanced variant of continuous batching~\cite{agrawal2023sarathi} attempts to balance TTFT and TPOT by segmenting long prefill into chunks and attaching decoding jobs with a chunked prefill  --  but essentially, it trades TTFT for TPOT and cannot eliminate the interference (\S\ref{sec:problem-opportunity}).
In summary, batching prefill and decoding invariably leads to compromises in either TTFT or TPOT.

\parabf{Model parallelism.} In LLM serving, model parallelism is generally divided as intra- and inter-operator parallelisms~\cite{li2023alpaserve,alpa,shoeybi2020megatronlm}. Both can be used to support larger models but may impact serving performance differently. 
Intra-operator parallelism partitions computationally intensive operators, such as matrix multiplications, across multiple GPUs, accelerating computation but causing substantial communication. 
It reduces the execution time\footnote{we emphasize ``execution time'' instead of latency here because latency comprises both execution time and queuing delay.}, hence latency, particularly for TTFT of the prefill phase, but requires high bandwidth connectivity between GPUs (e.g., NVLINK).
Inter-operator parallelism organizes LLM layers into stages, each running on a GPU to form pipelines. It moderately increases execution time due to inter-stage communication, but linearly scales the system's rate capacity with each added GPU.
In this paper, we reveal an additional benefit of model parallelism: reduced queuing delay of both prefill and decoding phases, steaming from shorter execution time. We delve into this further in \S\ref{sec:tradeoff}.
Besides model parallelism, replicating a model instance, irrespective of its model parallelism configurations, linearly scales the system's rate capacity.

These parallelism strategies create a complex space of optimization that requires careful trade-offs based on the application's latency requirements.

\subsection{Problems and Opportunities}
\label{sec:problem-opportunity}
Colocating and batching the prefill and decoding computation to maximize the overall system throughput, as in existing systems, is cost-effective for service providers. However, in the presence of SLOs, present approaches struggle to maintain both high service quality and low cost due to the issues discussed below.

\begin{figure}[t]
    \centering
    \includegraphics[width=\columnwidth]{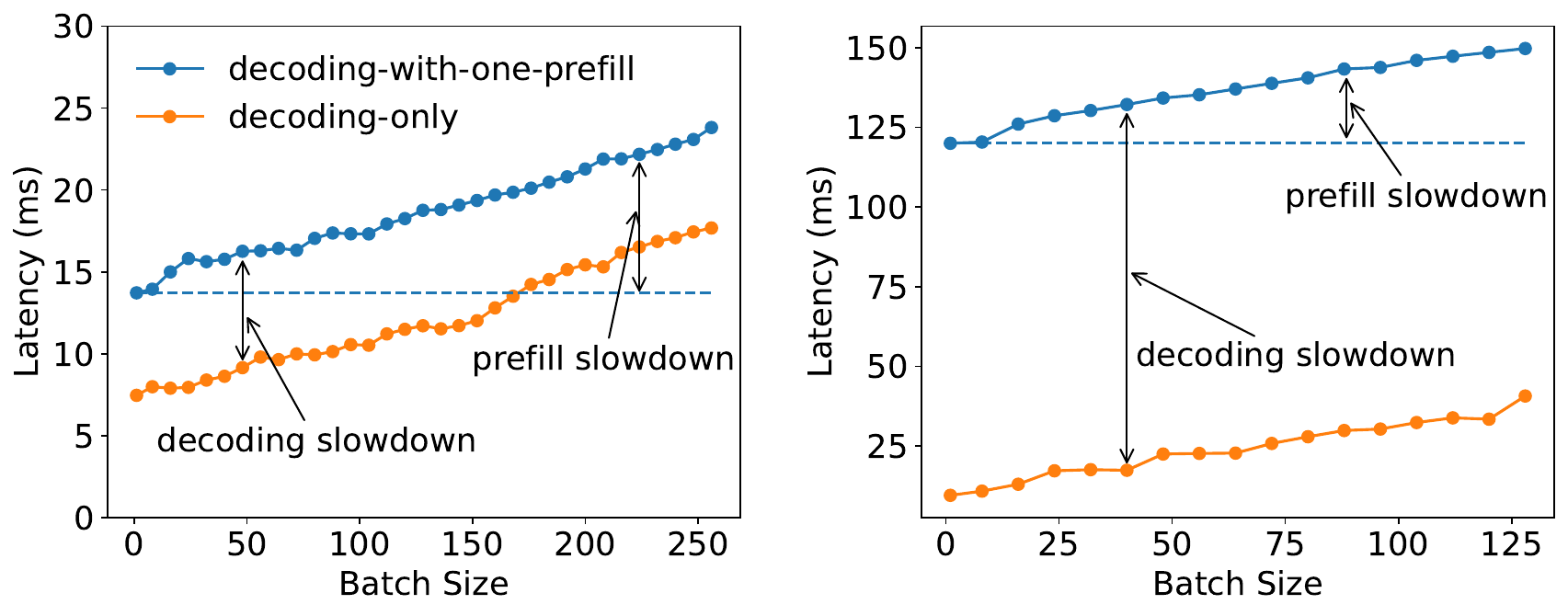}
    (a) Input length = 128 \hspace{11mm} (b) Input length = 1024
    \vspace{-2mm}
    \caption{Batch execution time when serving a 13B LLM as batch size increases. Compared between a decoding-only batch and the batch adding one more prefill job.}
    \vspace{-6mm}
    \label{fig:interference}
\end{figure}

\parabf{Prefill-decoding interference.} As Figure~\ref{fig:interference} shows, adding a single prefill job to a batch of decoding requests significantly slows down both processes, leading to a marked increase in TTFT and TPOT. 
Specifically, the decoding tasks in the batch must wait for lengthier prefill jobs to complete, thus extending TPOT; the slowdown intensifies with a longer prefill, shown in Figure~\ref{fig:interference}\textcolor{green!80!black}{(b)}.
Adding decoding jobs to prefill also increases the time to complete the prefill task, particularly when the GPU is already at capacity (Figure~\ref{fig:interference} blue curves). 

One attempt to mitigate this interference is called \textit{chunked-prefill with piggyback}~\cite{agrawal2023sarathi, deepspeed_mii}. It proposes to split the long prefill into chunks and batch a prefill chunk with a few decoding jobs (a.k.a. piggybacking). This technique alleviates the slowdown of the decoding job caused by the long prefill job, but it does not eliminate it. Additionally, it results in an extra overhead for the prefill job which cannot be easily mitigated by adjusting the chunk size. First, if the chunk size is set much lower than the inflection point that can saturate the GPU, then the prefill job will have a longer execution time since it competes with the decoding job in the same batch and cannot solely utilize the GPU resources. Second, if we increase the chunk size to nearly saturate the GPU, the chance of piggybacking will diminish since the remaining slots for decode tokens are limited. Also, chunked-prefill causes significantly more memory access for the prefill jobs. This is because the KV cache of all previous chunks have to be loaded from HBM to SRAM repeatedly to compute each subsequent chunk. Concretely, if a prefill job is split into $N$ equal chunks, we need to load $N + (N-1) + ... + 1 = O(N^2)$ chunks of KV Cache in total, compared to O(N) in the non-chunked case. This overhead will increase as the context length becomes longer.

\parabf{Ineffective scheduling.} Unbatching prefill and decoding jobs and scheduling them sequentially does not mitigate the interference. Decoding jobs may experience longer queuing delays due to waiting for ongoing prefill jobs on GPUs. Moreover, batches dedicated to decoding often lead to GPU underutilization. Prioritizing tasks in either phase adversely affects the latency of the other, rendering priority scheduling ineffective.

\parabf{Resource and parallelism coupling.} Colocating prefill and decoding phases on the same GPUs unavoidably share their resource and parallelism settings. However, each phase has its unique computational characteristic and latency requirement that calls for more heterogeneous resource allocation. 
For example, the prefill phase tends to be compute-bound and benefits from more intra-op parallelism to reduce execution time to meet the tight SLO on TTFT. By contrast, the optimal parallelism configuration of the decoding phase depends on the running batch size.
In existing systems, due to coupling, resource allocation and parallelism plans are tailored to satisfy the \emph{more demanding} of TTFT and TPOT, which may not be ideal for the other. This often leads to resource over-provisioning to meet both SLOs. 

\parabf{Opportunities.} To address these issues, we propose to disaggregate the prefill and decoding phases. We use the term \textit{instance} to denote a unit of resources that manages exactly one complete copy of model weights. One instance can correspond to many GPUs when model parallelism is applied.
Note that when we disaggregate the two phases to different GPUs, each phase manages its copy of the model weights, resulting in \emph{prefill instances} and \emph{decoding instances}. A prefill instance, upon receiving a request, performs only the prefill computation for this request to generate the first output token. It then sends the intermediate results (mainly KV caches) to a decoding instance, which is responsible for subsequent decoding steps. 
Because decoding computation often has low GPU utilization, we may allocate multiple prefill instances per decoding instance. This allows batching more decoding jobs to achieve higher GPU utilization.

Disaggregating prefill and decoding naturally resolves the interference between the two phases and enables each to focus on its optimization target -- TTFT or TPOT. Each type of instance can employ different resources and parallelism strategies to meet a variety of latency requirements. 
By adjusting the number of GPUs and parallelisms provided to the two types of instances, we can maximize the per-device goodput of the overall system, avoiding over-provisioning, eventually translating to reduced cost-per-query adhering to service quality. Next, we develop ways to find out the best resource allocation and parallelism plan for each phase.
\section{Tradeoff Analysis}
\label{sec:tradeoff}
Disaggregation uncouples the two phases and allows a distinct analysis of the characteristics of each phase, providing valuable insights into the algorithm design. It also expands the design space: now each phase needs to be scaled and scheduled independently based on their latency requirements.

In this section, we analyze the computational pattern of prefill (\S\ref{subsec:analysis_prefill}) and decoding instances (\S\ref{subsec:analysis_decoding}) \emph{post disaggregation}. We aim to identify key parameters and derive guidelines for batching and parallelism in each phase. We then highlight several practical deployment considerations (\S\ref{subsec:practical_problems}). This section lays the foundation for per-gpu goodput optimization.

\subsection{Analysis for Prefill Instance}
\label{subsec:analysis_prefill}

\begin{figure}[!t]
    \centering
    \includegraphics[width=\columnwidth]{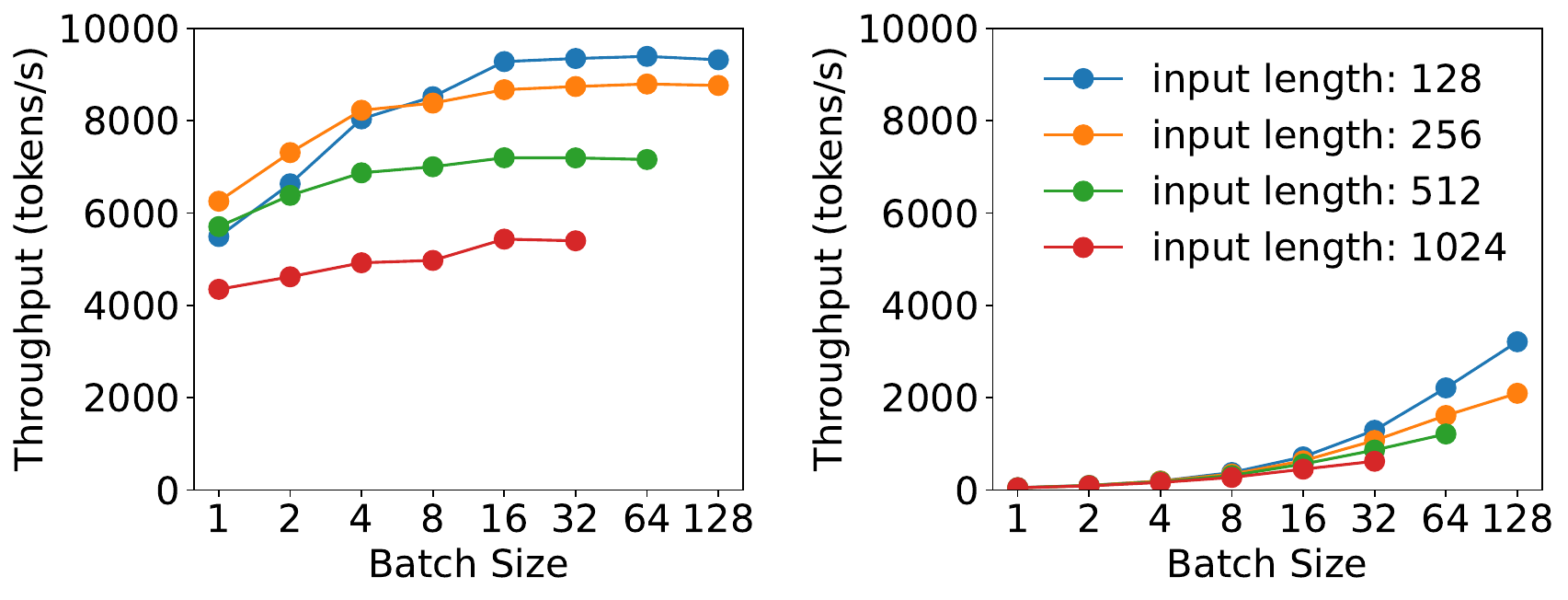}
    (a) Prefill phase \hspace{50pt} (b) Decoding phase \hspace{-30pt} 
    \caption{Throughput for two phases with different batch sizes and input lengths when serving an LLM with 13B parameters.}
    \vspace{-4mm}
    \label{fig:prefill_decode_tpt}
\end{figure}

After disaggregation, the prefill phase generates the first token by processing all tokens of the user prompt in parallel. Assuming a given arrival rate, we aim to fulfill the service's latency requirement on TTFT using the least resources.

\parabf{Batching strategy.} The prefill step is typically compute-intensive. Figure~\ref{fig:prefill_decode_tpt}\textcolor{green!80!black}{(a)} shows how the throughput of the prefill phase changes with the input length and the batch size. For a 13B parameter LLM, processing a single sequence of 512 tokens can fully engage an A100 GPU. 
Once the GPU becomes compute-bound, adding more requests to the batch no longer improves GPU efficiency. Instead, it proportionally extends the total processing time for the batch, inadvertently delaying all included requests. Hence, for prefill instances, it is necessary to profile the specific LLM and GPUs in advance to identify a critical input length threshold, denoted as $L_m$, beyond which the prefill phase becomes compute-bound. Batching more requests should only be considered when the input length of the scheduled request is below $L_m$. 
In practice, user prompts typically average over hundreds of tokens~\cite{sharegpt}. Batch sizes for the prefill instance are generally kept small.

\begin{figure}[!t]
    \centering
    \includegraphics[width=\columnwidth]{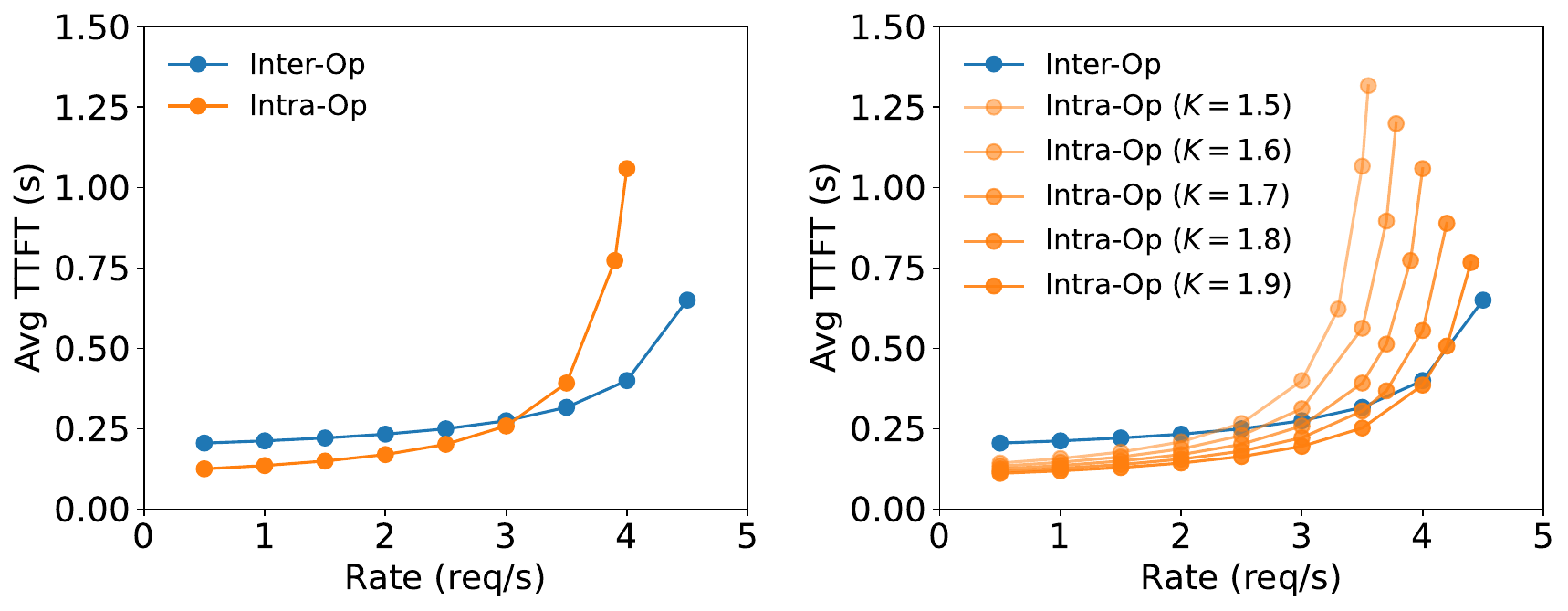}
    (a) Real experiment results \hspace{7pt} (b) Changing intra-op speedup
    \vspace{-0.5mm}
    \caption{Average TTFT when serving an LLM with 66B parameters using different parallelism on two A100 GPUs. }
    \vspace{-4.9mm}
    \label{fig:prefill_intra_op}
\end{figure}

\parabf{Parallelism plan.} 
To study the parallelism preferences for prefill-only instances, we serve a 66B LLM on two A100 GPUs with inter-op or intra-op parallelism strategy. To simplify the problem, we assume uniform requests input lengths of 512 tokens and a Poisson arrival process. We compare the resulting average TTFT at various arrival rates in Figure~\ref{fig:prefill_intra_op}\textcolor{green!80!black}{(a)}: intra-op parallelism is more efficient at lower arrival rates, while inter-op parallelism gains superiority as the rate increases.
Disaggregation enables the prefill phase to function analogously to an M/D/1 queue, so we can use queuing theory to verify the observation.

We start by developing notations using the single-device case without parallelism: 
each request's execution time, denoted as $D$, remains constant due to uniform prefill length. Since one request saturates the GPU, we schedule requests via First-Come-First-Served (FCFS) without batching. 
Suppose the Poisson arrival rate is $R$ and the utilization condition of $RD < 1$, the average TTFT ($Avg\_TTFT$) can be modeled by the M/D/1 queue~\cite{shortle2018fundamentals} in close form:
\begin{equation}
\small
    Avg\_TTFT = D + \frac{RD^2}{2(1-RD)}.
    \label{eq1}
\end{equation}
where the first term represents the execution time and the second corresponds to the queuing delay. 
Based on Eq.~\ref{eq1}, we incorporate parallelism below. 

With 2-way inter-op parallelism, we assume the request-level latency becomes $D_s$, and the slowest stage takes $D_m$ to finish. We have $D \approx D_s \approx 2 \times D_m$, due to negligible inter-layer activation communication~\cite{alpa,li2023alpaserve}. The average TTFT with 2-way inter-op parallelism is derived as:

\begin{equation}
\small
    Avg\_TTFT_{inter} = D_s + \frac{RD_m^2}{2(1-RD_m)} = D + \frac{RD^2}{4(2-RD)}.
    \label{eq2}
\end{equation}

For intra-op parallelism, we introduce a speedup coefficient $K$, where $1 < K < 2$, reflecting the imperfect speedup caused by high communication overheads of intra-op parallelism. 
With the execution time $D_s = \frac{D}{K}$, the average TTFT for 2-degree intra-op parallelism is:

\begin{equation}
\small
    Avg\_TTFT_{intra} = \frac{D}{K} + \frac{RD^2}{2K(K-RD)}.
    \label{eq3}
\end{equation}

Comparing Eq.~\ref{eq2} and Eq.~\ref{eq3}: at lower rates, where execution time (first term) is the primary factor, intra-op parallelism's reduction in execution time makes it more efficient. As the rate increases and the queuing delay (second term) becomes more significant, inter-op parallelism becomes advantageous, concurred with Figure~\ref{fig:prefill_intra_op}\textcolor{green!80!black}{(a)}.

The prefill phase's preference for parallelism is also influenced by TTFT SLO and the speedup coefficient $K$. 
Seen from Figure~\ref{fig:prefill_intra_op}\textcolor{green!80!black}{(a)}: A more stringent SLO will make intra-op parallelism more advantageous, due to its ability to reduce execution time.
The value of K depends on factors such as the input length, model architecture, communication bandwidth, and placement~\cite{shoeybi2020megatronlm,alpa}. As shown in Figure~\ref{fig:prefill_intra_op}\textcolor{green!80!black}{(b)}, a decrease in K notably reduces the efficacy of intra-op parallelism.
\S\ref{sec:method} develops algorithms that optimize the resource and parallelism configurations taking into consideration these knobs. 

\subsection{Analysis for Decoding Instance}
\label{subsec:analysis_decoding}
Unlike the prefill instance, a decoding instance follows a distinct computational pattern: it receives the KV caches and the first output token from the prefill instance and generates subsequent tokens one at a time. 
For decoding instances, our optimization goal is to satisfy the application's TPOT requirement using minimal computing resources.

\parabf{Batching strategy.} Since a single decoding job is heavily bandwidth-bound, batching is key to avoiding low GPU utilization (hence high per-gpu goodput), as shown in Figure~\ref{fig:prefill_decode_tpt}\textcolor{green!80!black}{(b)}. 
In existing systems where the prefill and decoding phases are colocated, increasing the decoding batch size is difficult because it conflicts with meeting latency goals, particularly in scenarios with high request rates. This is because sharing GPUs cause competition between prefill and decoding jobs, leading to a trade-off between TTFT and TPOT. For example, a higher arrival rate generates more prefill jobs, demanding greater GPU time to meet TTFT requirements if prioritizing prefill jobs, which in turn adversely affects TPOT.

On the contrary, disaggregation offers a solution by enabling the allocation of multiple prefill instances to a single decoding instance. This approach allows for accumulating a larger batch size on dedicated GPUs for the decoding phase without sacrificing TPOT.

\begin{figure}[!t]
    \centering
    \includegraphics[width=\columnwidth]{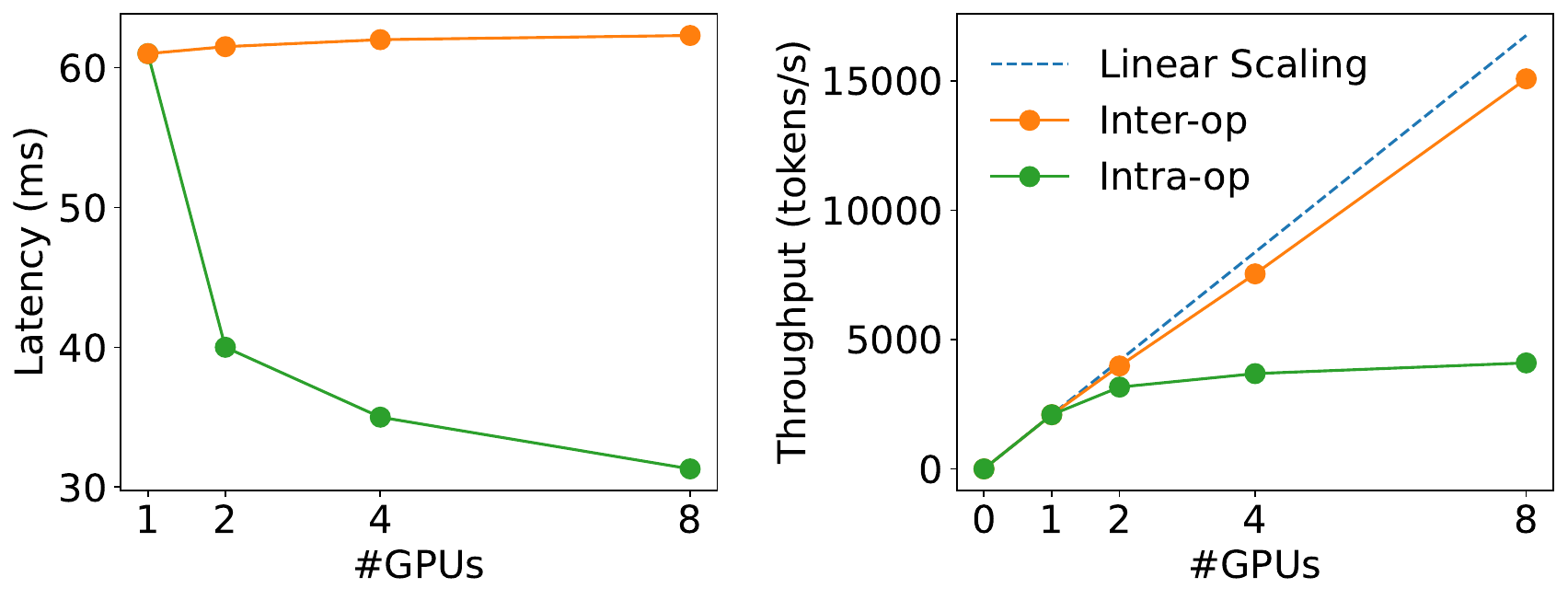}
    \caption{Decoding phase latency and throughput when serving a 13B LLM with batch size = 128 and input length = 256 under different parallel degrees.}
    \vspace{-4mm}
    \label{fig:decode_parallel}
\end{figure}

\parabf{Parallelism plan.} Post-disaggregation, the batch size for decoding may be constrained by GPU memory capacity, as it is necessary to maintain the KV caches for all active requests.
Scaling the decoding instance with model parallelism or leveraging advanced memory management techniques for LLM KV caches, such as Paged-Attention~\cite{vllm} and GQA~\cite{ainslie2023gqa}, enable further scaling of the decoding batch size to nearly compute-bound. 
As the decoding batch size continue to increase to approach the compute-bound, the decoding computation begins to resemble the prefill phase. With this observation, we investigate how the latency and throughput change under different parallelism degrees under large batch conditions in Figure~\ref{fig:decode_parallel}: intra-op parallelism reduces latency with diminishing returns, caused by communication and reduced utilization after partitioning. Inter-op parallelism can almost linearly scale the throughput. Hence, when the TPOT SLO is stringent, intra-op parallelism is essential to reduce TPOT to meet latency goals. Beyond this, inter-op parallelism is preferable to enhance throughput linearly.

It is worth noting that when the model can fit into the memory of a single GPU, replication is a competitive option in addition to model parallelism for both prefill and decoding instances, to linearly scale the system's rate capacity. It may also reduce the queuing delay -- as indicated by Eq.~\ref{eq1}  -- by substituting $R$ with $R/N$ assuming requests are equally dispatched to $N$ replicas, at the cost of maintaining additional replicas of the model weights in GPU memory.

\subsection{Practical Problems}
\label{subsec:practical_problems}
We have developed foundational principles for selecting batching and parallelisms for each phase. In this section, we discuss and address several challenges encountered during the practical deployment of disaggregated prefill and decoding phases.

\parabf{Variable prefill length.} \S\ref{sec:tradeoff} has assumed uniform prompt length across requests. In real deployments, depending on the LLM application, the lengths of requests are non-uniform. The non-uniformity can cause pipeline bubbles~\cite{huang2019gpipe,pipedream} for prefill instances applying inter-op parallelism because the execution time of pipeline stages across requests of different lengths will vary. This results in slight deviations from the conclusions indicated by using the M/D/1 queue model. To address the problem, \S\ref{sec:method} develops algorithms that search for parallelisms based on workloads, and resort to scheduling to minimize the bubbles (\S\ref{subsec:online_scheduling}).

\parabf{Communication overhead.}
Transferring KV caches from prefill to decoding instances incurs notable overheads. For example, the KV cache size of a single 512-token request on OPT-66B is approximately 1.13GB. Assuming an average arrival rate of 10 rps, we need to transfer 11.3GB data per second---or equivalently 90Gbps bandwidth to render the overhead invisible. While many modern GPU clusters for LLMs are equipped with InfiniBand (e.g., 800 Gbps), in cases where cross-node bandwidth is limited, \sys relies on the commonly available intra-node NVLINK, where the peak bandwidth between A100 GPUs is 600 GB/s, again rendering the transmission overhead negligible (see \S\ref{exp:breakdown}). However, this requirement imposes additional constraints on the placement of prefill and decoding instances that we take into consideration in the next section.

Through the analysis in this section, we identify the workload pattern, placement constraints, SLO requirements, parallelism strategies, and resource allocation as key parameters that create a web of considerations in designing the disaggregated serving system. How to automatically navigate the search space to find the configuration that achieves optimal per-gpu goodput is challenging, and addressed next.

\section{Method}
\label{sec:method}

We built \sys to solve the above challenges. Given the model, workload characteristic, latency requirements, and SLO attainment target, \sys will determine (a) the parallelism strategies for prefill and decoding instances, (b) the number of each instance type to deploy, as well as (c) how to place them onto the physical cluster. We call the solution a \textit{placement}. Our goal is to find a placement that maximizes the per-gpu goodput.

As explained in \S\ref{subsec:practical_problems}, a key design consideration is to manage communications between disaggregated prefill and decoding phases, given varying cluster setups.
In this section, we first present two placement algorithms: one for clusters with high-speed cross-node networks (\S\ref{subsec:optimal_placement}) and the other for environments lacking such infrastructure (\S\ref{subsec:practical_placement}); the latter introduces additional constraints. We then develop online scheduling optimizations that adapt to the nuances of real-world workloads (\S\ref{subsec:online_scheduling}).

\subsection{Placement for High Node-Affinity Cluster}
\label{subsec:optimal_placement}

\begin{algorithm}[t!]
\caption{High Node-Affinity Placement Algorithm}\label{alg:optimal-placement}
\begin{algorithmic}
\Require LLM $G$, \#node limit per-instance $N$, \#GPU per-node $M$, GPU memory capacity $C$, workload $W$, traffic rate $R$.
\Ensure the placement $\mathit{best\_plm}.$ 
    \State $\mathit{config_p}, \mathit{config_d} \leftarrow \emptyset, \emptyset$
    \For{$\mathit{intra\_op} \in \{1, 2, ..., M\}$}
        \For{$\mathit{inter\_op} \in \{1, 2, ..., \frac{N\times M}{\mathit{intra\_op}}\}$}
            \If{$\frac{G.size}{\mathit{inter\_op} \times \mathit{intra\_op}} < C$}
                \State $\mathit{config} \leftarrow (\mathit{inter\_op}, \mathit{intra\_op})$
                \State $\hat{G} \leftarrow \text{parallel}(G, \mathit{config})$
                \State $\mathit{config.goodput} \leftarrow \text{simu\_prefill}(\hat{G}, W)$
                \If{$\frac{\mathit{config_p.goodput}}{config_p.num\_gpus} < \frac{\mathit{config.goodput}}{config.num\_gpus}$}
                    \State $\mathit{config_p} \leftarrow \mathit{config}$
                \EndIf
                \State $\mathit{config.goodput} \leftarrow \text{simu\_decode}(\hat{G}, W)$
                \If{$\frac{\mathit{config_d.goodput}}{config_d.num\_gpus} < \frac{\mathit{config.goodput}}{config.num\_gpus}$}
                    \State $\mathit{config_d} \leftarrow \mathit{config}$
                \EndIf
            \EndIf
        \EndFor
    \EndFor
    \State $n, m \leftarrow \lceil \frac{R}{\mathit{config_p.goodput}} \rceil, \lceil \frac{R}{\mathit{config_d.goodput}} \rceil$ 
    \State $\mathit{best\_plm} \leftarrow (n, \mathit{config_p}, m, \mathit{config_d})$
\State \textbf{return} $\mathit{best\_plm}$
\end{algorithmic}
\end{algorithm}

On high node-affinity clusters equipped with Infiniband, KV caches transmission overhead across nodes is negligible, \sys can deploy prefill and decoding instances across any two nodes without constraints. 
We propose a two-level placement algorithm for such scenarios:  we first optimize the parallelism configurations for prefill and decoding instances separately to attain phase-level optimal per-gpu goodput; then, we use replication to match the overall traffic rate.

However, finding the optimal parallel configuration for a single instance type, such as for the prefill instance, is still challenging, due to the lack of a simple analytical formula to calculate the SLO attainment (a.k.a., percentage of requests that meet TTFT requirement), given that the workload has diverse input, output lengths, and irregular arrival patterns. Gauging the SLO via real-testbed profiling is time-prohibitive. We thus resort to building a simulator to estimate the SLO attainment, assuming prior knowledge of the workload's arrival process and input and output length distributions.
Although short-term interval is impossible to predict, the workload pattern over longer timescales (e.g.,
hours or days) is often predictable~\cite{li2023alpaserve, zhangshepherd}. \sys fits a distribution from the history request traces and resamples new traces from the distribution as the input workload to the simulator to compute the SLO attainment. Next, \sys simply enumerates the placements and finds the maximum rate that meets the SLO attainment target with binary search and simulation trials.

Algorithm~\ref{alg:optimal-placement} outlines the process. We enumerate all feasible parallel configurations, subject to cluster capacity limit, for both prefill and decoding instances. Then, for a specific prefill phase configuration, we use \verb|simu_prefill| to simulate and find its maximum goodput via binary search (similarly for using \verb|simu_decode|  for decoding). 
After determining the optimal parallel configurations for both prefill and decoding instances, we replicate them to achieve the user-required overall traffic rate according to their goodput. 

The complexity of Algorithm~\ref{alg:optimal-placement} is $O(NM^2)$, with $N$ as the node limit per instance and $M$ representing the typical number of GPUs per node in modern clusters (e.g., 8). The search space is manageable and the solving time is under 1.3 minutes in our largest setting, as demonstrated in \S\ref{exp:run_time}.

\parabf{Simulator building.} Algorithm~\ref{alg:optimal-placement} relies on a simulator to estimate the goodput under various SLOs and SLO attainment goals given the workload and the parallelism plan. 
To build an accurate simulator, we analyze the FLOPs and the number of memory accesses for prefill and decoding phases respectively, and use a latency model to approximate the inference execution time. See details in Appendix~\ref{appendix:latency_model}. The simulator aligns well with real profiling results, thanks to the high predictability of DNN workloads~\cite{clockwork, li2023alpaserve}, verified in \S\ref{exp:ablation}. 

By far, we have developed Algorithm~\ref{alg:optimal-placement} assuming we can place the prefill and decoding instance between any two nodes (or on the same node) of the cluster, and the KV cache transmission utilizes high bandwidth network. In many real clusters, GPUs inside a node access to high-bandwidth NVLINK while GPUs distributed across nodes have limited bandwidth. We next develop an algorithm to address this constraint.

\begin{algorithm}[t!]
\caption{Low Node-Affinity Placement Algorithm}\label{alg:low-affinity-placement}
\begin{algorithmic}
\Require LLM $G$, \#node limit per-instance $N$, \#GPU per-node $M$, GPU memory capacity $C$, workload $W$, traffic rate $R$.
\Ensure the placement $\mathit{best\_plm}.$ 
    \State $\mathit{config}^{*} \leftarrow \emptyset$
    \For{$\mathit{inter\_op} \in \{1, 2, ..., N\}$}
        \State $\mathcal{P} \leftarrow \text{get\_intra\_node\_configs}(G, M, C, \mathit{inter\_op})$
        \For{$P_p \in \mathcal{P}$}
            \For{$P_d \in \mathcal{P}$}
                \If{$P_p.\mathit{num\_gpus} + P_d.\mathit{num\_gpus} \le M$}
                    \State $\mathit{config} \leftarrow (\mathit{inter\_op}, P_p, P_d)$
                    \State $\hat{G}_p, \hat{G}_d \leftarrow \text{parallel}(G, \mathit{config})$
                    \State $\mathit{config.goodput} \leftarrow \text{simulate}(\hat{G}_p, \hat{G}_d, W)$
                        \If{$\frac{\mathit{config.^{*}goodput}}{\mathit{config.^{*}num\_gpus}} < \frac{\mathit{config.goodput}}{\mathit{config.num\_gpus}}$}
                            \State $\mathit{config^{*}} \leftarrow \mathit{config}$
                        \EndIf
                \EndIf
            \EndFor
        \EndFor
    \EndFor
    
    \State $n \leftarrow \lceil \frac{R}{\mathit{config.^{*}goodput}} \rceil$ 
    \State $\mathit{best\_plm} \leftarrow (n, \mathit{config^{*}})$

\State \textbf{return} $\mathit{best\_plm}$
\end{algorithmic}
\end{algorithm}

\subsection{Placement for Low Node-Affinity Cluster}
\label{subsec:practical_placement}
A straightforward solution is to always colocate prefill and decoding instances on the same node, utilizing the NVLINK, which is commonly available inside a GPU node.
For large models, e.g. with 175B parameters (350GB), we may be unable to even host a single pair of prefill and decoding instances in an 8-GPU node ($80G \times 8 = 640G < 350 \times 2 GB $). We incorporate this as additional placement constraints and co-optimize it with model parallelism, presented in Algorithm~\ref{alg:low-affinity-placement}.

The key insight is that KV cache transfer occurs exclusively between corresponding layers of prefill and decoding instances.
Leveraging inter-op parallelism, we group layers into stages and divide each instance into segments, termed as \textit{instance segments}, with each segment maintaining one specific inter-op stage. 
By colocating prefill and decoding segments of the same stage within a single node, we force the transfer of intermediate states to occur only via NVLINK. Inside a node, we set the same parallelism and resource allocation for segments of the same instance. Given the typical limitation of GPUs per node (usually 8), we can enumerate possible configurations inside one node and use the simulator to identify the configurations that yield the best goodput.

As outlined in Algorithm~\ref{alg:low-affinity-placement}, we begin by enumerating inter-op parallelism degrees to get all the possible instance segments. For each segment, we get all possible intra-node parallelism configurations by calling \verb|get_intra_node_configs|. Then we use simulation to find the optimal one and replicate it to satisfy the target traffic rate.

\subsection{Online scheduling}
\label{subsec:online_scheduling}

\begin{figure}[!t]
    \centering
    \includegraphics[width=0.9\linewidth]{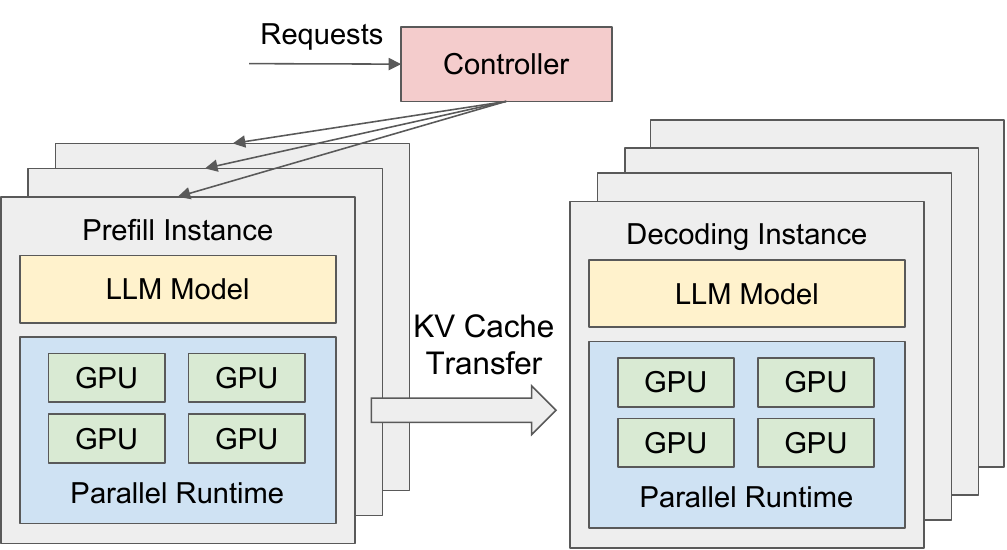}
    \caption{\sys Runtime System Architecture}
    \vspace{-4mm}
    \label{fig:architecture}
\end{figure}

The runtime architecture of \sys is shown in Figure~\ref{fig:architecture}. \sys operates with a simple FCFS scheduling policy. All incoming requests arrive at a centralized controller, then dispatched to the prefill instance with the shortest queue for prefill processing, followed by dispatch to the least loaded decoding instance for decoding steps. This setup, while simple, is optimized with several key enhancements tailored to the nuances of real-world workloads.

\parabf{Reducing pipeline bubbles.} 
To mitigate the pipeline bubbles caused by non-uniform prompt lengths (\S\ref{subsec:practical_problems}), we schedule the requests in a way that balances the execution time across all batches in the pipeline. This is achieved by noting that, for both prefill and decoding instances, the number of new tokens in the batch is a reliable indicator of the batch's real execution time.
For prefill instances, we profile the target model and GPU to figure out the shortest prompt length $L_m$ needed to saturate the GPU. We schedule prefill batches with a total sequence length close to $L_m$, by either batching multiple requests shorter than $L_m$ or individually scheduling requests longer than $L_m$. For decoding instances, we set $L_m$ as the largest batch size.

\parabf{Combat busrtiness.} 
Burstiness in workloads can cause a deluge of KV caches to transfer from prefill to decoding instances, risking memory overload on decoding instances.
To circumvent this, \sys employs a ``pull'' method for KV cache transmission rather than a ``push'' approach -- decoding instances fetch KV cache from prefill instances \emph{as needed}, using the GPU memory of prefill instances as a queuing buffer. This way, the prefill instance can continue handling other prefill jobs by simply retaining the KV Cache in the GPU memory after processing the prompt. Hence, each type of instance operates at its own pace without complex coordination.

\parabf{Replaning.} The resource and parallelism plan in \sys is optimized for a specific workload pattern, which may become suboptimal if the workload pattern changes over time. \sys implement periodic replanning. A workload profiler monitors key parameters such as the average input and output length of the requests, the average arrival rate, etc. If a significant pattern shift is detected, \sys will trigger a rerun of the placement algorithm based on recent historical data. This process is expedient -- the proposed algorithm runs in seconds (\S\ref{exp:run_time}) and reloading LLM weights can be completed within minutes -- far shorter than the hourly scale at which real-world workload variations tend to occur.

\parabf{Preemption and fault tolerance.} \sys does not implement advanced runtime policies like preemption~\cite{preemption} and fault tolerance~\cite{fault_tolerance}, which are complementary to disaggregation. Nevertheless, we discuss how they fit into \sys.
In \sys, the FCFS policy can lead to a ``convoy effect'', where longer requests block shorter ones in the prefill stage. Incorporating preemptive strategies, as suggested in existing literature \cite{wu2023fast}, could enhance efficiency and is feasible within our system's architecture.
While not a primary focus in the current \sys, fault tolerance is a critical aspect for consideration. In traditional colocation- and replication-based systems, a fault in one instance typically does not disrupt other replica instances. However, in \sys, the dependency between prefill and decoding instances introduces the risk of fault propagation. For example, a fault in a single decoding instance mapped to multiple prefill instances could potentially cripple the entire service and cluster. We leave both as future work.

\section{Implementation}
\label{sec:implementation}

\sys is an end-to-end distributed serving system for LLMs with a placement algorithm module, a RESTful API frontend, an orchestration layer, and a parallel execution engine. The algorithm module, frontend, and orchestration layer are implemented with 6.5K lines of Python code. The parallel execution engine is implemented with 8.1K lines of C++/CUDA code.

The placement algorithm module implements the algorithm and the simulator mentioned in \S\ref{sec:method} which gives the placement decision for a specific model and cluster setting. The frontend supports an OpenAI API-compatible interface where clients can specify the sampling parameters like maximum output length and temperature. The orchestration layer manages the prefill and decoding instances, responsible for request dispatching, KV cache transmission, and results delivery. It utilizes NCCL~\cite{nccl} for cross-node GPU communication and asynchronous CudaMemcpy for intra-node communication, which avoids blocking the GPU computation during transmission. Each instance is powered by a parallel execution engine, which uses Ray~\cite{ray} actor to implement GPU workers that execute the LLM inference and manage the KV Cache in a distributed manner. It integrates many recent LLM optimizations like continuous batching~\cite{yu2022orca}, FlashAttention~\cite{dao2022flashattention}, PagedAttention~\cite{vllm} and supports popular open-source LLMs such as OPT~\cite{zhang2022opt} and LLaMA~\cite{touvron2023llama}.

\section{Evaluation}
\label{sec:evaluation}
In this section, we evaluate \sys under different sizes of LLMs ranging from 13B to 175B and various application datasets including chatbot, code-completion, and summarization. The evaluation shows that \sys consistently outperforms the current state-of-the-art system across all the settings (\S\ref{exp:end_to_end}). Specifically, \sys can handle up to $7.4\times$ higher rates and $12.6\times$ more stringent SLO while meeting the latency requirements for over 90\% requests. Additionally, we analyze the latency breakdown in \sys to show the communication overhead is insubstantial thanks to our bandwidth-aware placement algorithm (\S\ref{exp:breakdown}) and do ablation studies of our techniques (\S\ref{exp:ablation}). Finally, we profile the execution time of our placement algorithm (\S\ref{exp:run_time}).

\begin{table}[t]
    \centering
    \resizebox{1\linewidth}{!} {
    \begin{tabular}{ccccc}
        \toprule
        \textbf{Application} & \textbf{Model Size} & \textbf{TTFT} & \textbf{TPOT} & \textbf{Dataset} \\
        \midrule
        Chatbot OPT-13B & 26GB & 0.25s & 0.1s & ShareGPT~\cite{sharegpt} \\
        Chatbot OPT-66B & 132GB & 2.5s & 0.15s & ShareGPT~\cite{sharegpt} \\
        Chatbot OPT-175B & 350GB & 4.0s & 0.2s & ShareGPT~\cite{sharegpt} \\
        Code Completion OPT-66B & 132GB & 0.125s & 0.2s & HumanEval~\cite{chen2021codex} \\
        Summarization OPT-66B & 132GB & 15s & 0.15s & LongBench~\cite{bai2023longbench} \\
        \bottomrule
    \end{tabular}
    }
    \caption{Workloads in evaluation and latency requirements.}
    \label{tab:app_config}
\end{table}

\begin{figure}[!t]
    \centering
    \includegraphics[width=\linewidth]{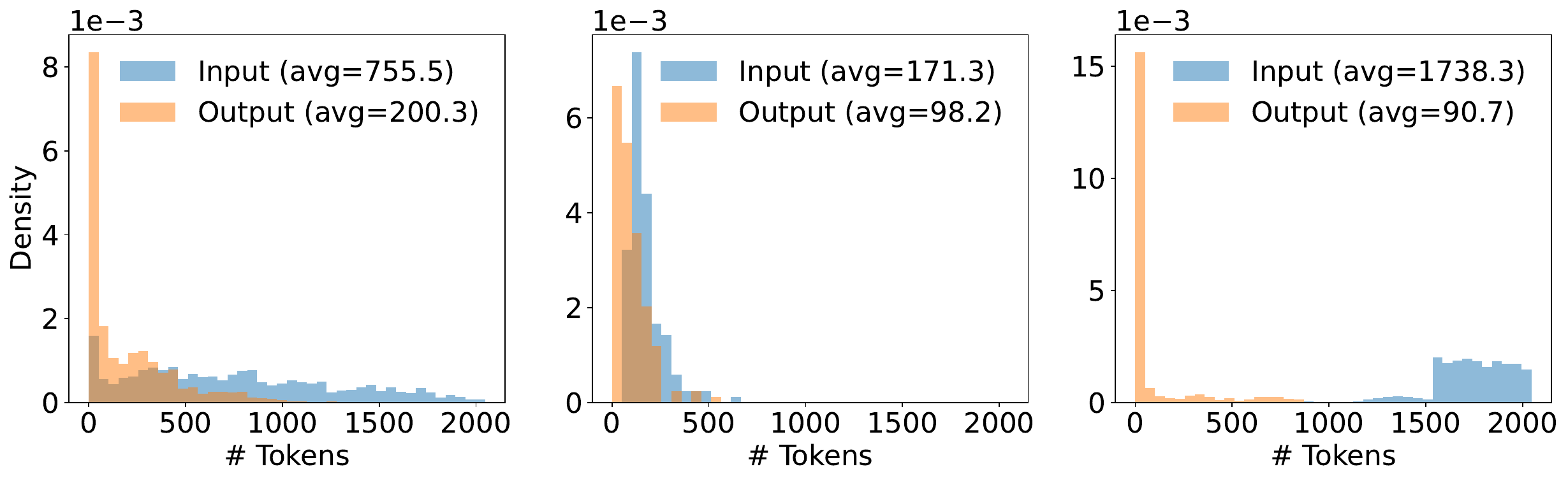}
    (a) ShareGPT \hspace{8mm} (b) HumanEval \hspace{8mm} (c) LongBench
    \caption{The input and output length distributions of (a) ShareGPT, (b) HumanEval, and (c) LongBench datasets.}
    \vspace{-4mm}
    \label{fig:dataset}
\end{figure}

\begin{figure*}[!t]
    \centering
    \includegraphics[width=0.95\linewidth]{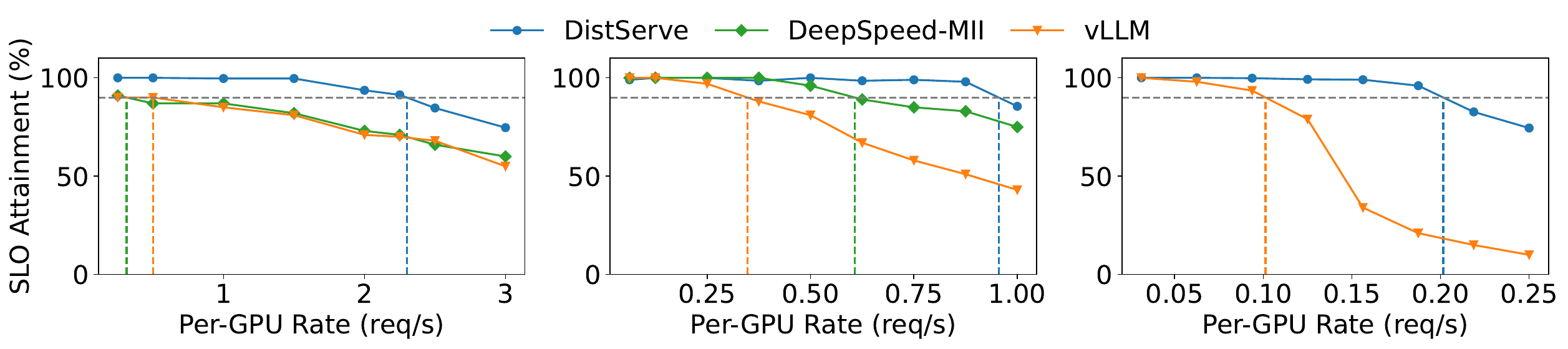}
    \includegraphics[width=0.95\linewidth]{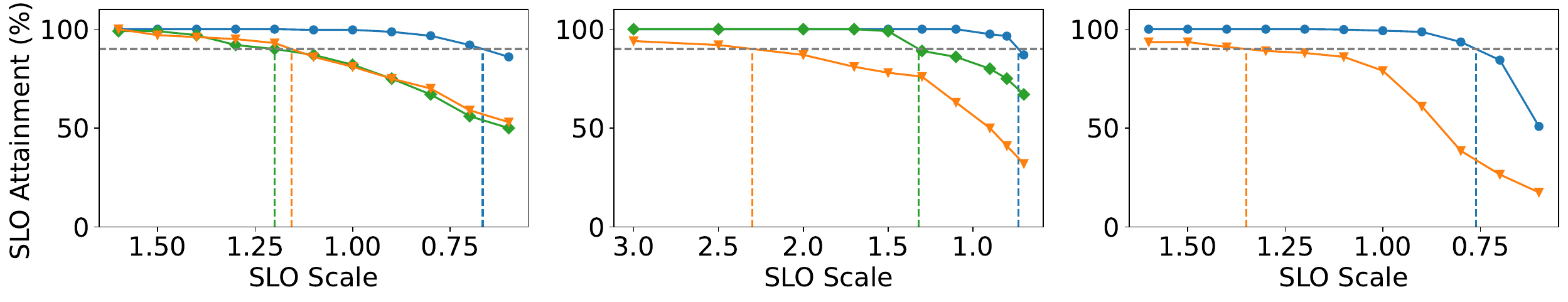}
    \hspace{10mm} (a) OPT-13B \hspace{40mm} (b) OPT-66B \hspace{40mm} (C) OPT-175B
    \caption{Chatbot application with OPT models on the ShareGPT dataset.}
    \vspace{-4mm}
    \label{fig:chatbot}
\end{figure*}

\subsection{Experiments Setup}
\paraf{Cluster testbed.} We deploy \sys on a cluster with 4 nodes and 32 GPUs. Each node has 8 NVIDIA SXM A100-80GB GPUs connected with NVLINK. The cross-node bandwidth is 25Gbps. Due to the limited cross-node bandwidth, we use the low node-affinity placement algorithm (\S\ref{alg:low-affinity-placement}) for \sys in most of the experiments except for the ablation study (\S\ref{exp:ablation}) which uses simulation.

\parabf{Model and workloads setup.} Similar to prior work on LLM serving~\cite{vllm}, we choose the OPT~\cite{zhang2022opt} model series, which is a representative LLM family widely used in academia and industry. Newer GPT model families are adopting memory-efficient attention mechanisms like GQA~\cite{ainslie2023gqa} and MQA~\cite{mqa}. \sys will show better performance on these models because the transmission overhead is lower due to the decrease in KV cache size. We choose OPT which uses the classic MHA~\cite{vaswani2017attention} to put enough pressure on the transmission overhead. We use FP16 precision in all experiments. For workloads, as shown in Table~\ref{tab:app_config}, We choose three typical LLM applications and set the SLOs empirically based on their service target because there exists no available SLO settings for these applications as far as we know. For each application, we select a suitable dataset and sample requests from it for evaluation. Since all the datasets do not include timestamps, we generate request arrival times using Poisson distribution with different request rates. Due to the space limit, we test the chatbot workload on all three OPT models and the other two workloads on OPT-66B, which matches the largest size in the recent open-source LLM series~\cite{touvron2023llama}.
\begin{itemize}[leftmargin=*]
    \item \textbf{Chatbot}~\cite{chatgpt}: We use the ShareGPT dataset~\cite{sharegpt} for the chatbot application, which is a collection of user-shared conversations with ChatGPT. For OPT-13B, the TTFT SLO is set to 0.25s for responsiveness and the TPOT SLO is set to 0.1s which is higher than the normal human read speed. For OPT-66B and OPT-175B, we slightly relax the two SLOs due to the increase in model execution latency. 
    \item \textbf{Code completion}~\cite{chen2021codex}: We use the HumanEval~\cite{chen2021codex} dataset for the code completion task. It includes 164 programming problems with a function signature or docstring which is used to evaluate the performance of code completion models. Since the code completion model is used as a personal real-time coding assistant, we set both SLOs to be stringent.
    \item \textbf{Summarization}~\cite{lanchain}: It is a popular LLM task to generate a concise summary for a long article, essay, or even an academic paper. We use LongBench~\cite{bai2023longbench} dataset which contains the summarization task\footnote{We capped the input lengths in LongBench because OPT’s absolute positional embedding only supports a maximum length of 2048.}. As shown in Figure~\ref{fig:dataset}, LongBench has much longer input lengths than the other two datasets. So we set a loose TTFT SLO but require a stringent TPOT.
\end{itemize}

\parabf{Metrics.} We use \textit{SLO attainment} as the major evaluation metric. Under a specific SLO attainment goal (say, 90\%), we are concerned with two things: the maximum per-GPU goodput and the minimal SLO the system can handle. We are particularly interested in an SLO attainment of 90\% (indicated by the vertical lines in all curve plots), but will also vary the rate and latency requirements to observe how the SLO attainment changes. We also include the results in the Appendix for an SLO attainment of 99\% to show the system performance under a more stringent SLO attainment target.

\parabf{Baselines.} We compare \sys to two baseline systems:

\begin{itemize}[leftmargin=*]
    \item \textbf{vLLM~\cite{vllm}:} vLLM is a representative LLM serving system widely used in both academia and industry. It supports \textit{continuous batching}~\cite{yu2022orca} to increase throughput and \textit{paged-attention}~\cite{vllm} to reduce memory fragmentation during KV cache allocation. However, it colocates the prefill and decoding computation to maximize the overall system throughput and struggles to meet the latency requirements cost-efficiently. Since vLLM only supports intra-op parallelism, we follow previous work~\cite{vllm} to set intra-op equals 1, 4, and 8 for the three OPT models, respectively.
    \vspace{2mm}
    \item \textbf{DeepSpeed-MII~\cite{deepspeed_mii}:} DeepSpeed Model Implementations for Inference (MII) supports \textit{chunked-prefill} by decomposing long prompts into smaller chunks and composing with short prompts to exactly fill a target token budget. It mitigates but cannot eliminate the prefill-decoding interference caused by the long prefill job. We set its intra-op the same as vLLM for OPT-13B and OPT-66B for a fair comparison. However, DeepSpeed-MII cannot serve OPT-175B whose $vocab\_size=50272$ because its underlying kernel implementation requires $vocab\_size/intra\_op$ is a multiple of 8 where intra-op equals 8 does not satisfy. Setting intra-op equals 4 can satisfy this requirement but will cause the out-of-memory issue.
\end{itemize}

\begin{figure*}[!t]
    \centering
    \includegraphics[width=0.95\linewidth]{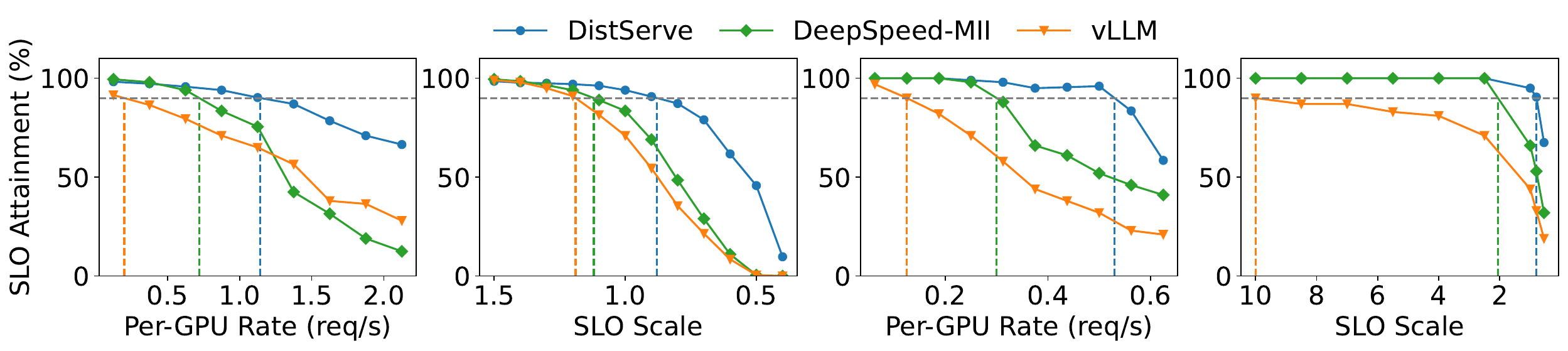}
    \hspace{10mm} (a) Code Completion \hspace{50mm} (b) Summarization
    \caption{Code completion and summarization tasks with OPT-66B on HumanEval and LongBench datasets, respectively.}
    \vspace{-4mm}
    \label{fig:other_apps}
\end{figure*}

\subsection{End-to-end Experiments}
\label{exp:end_to_end}
In this Section, we compare the end-to-end performance of \sys against the baselines on real application datasets.

\parabf{Chatbot.} We evaluate the performance of \sys on the chatbot application for all three OPT models. The first row of Figure~\ref{fig:chatbot} illustrates that when we gradually increase the rate, more requests will violate the latency requirements and the SLO attainment decreases. The vertical line shows the maximum per-GPU rate the system can handle to meet latency requirements for over 90\% of the requests.

On the ShareGPT dataset, \sys can sustain $2.0\times$--$4.6\times$ higher request rate compared to vLLM. This is because DistLLM eliminates the prefill-decoding interference through disaggregation. Two phases can optimize their own objectives by allocating different resources and employing tailored parallelism strategies. Specifically, by analyzing the chosen placement strategy\footnote{All the placements chosen by \sys can be found in Appendix~\ref{appendix:placement_strategies}.} for 175B, we find the prefill instance has inter-op = 3, intra-op = 3; and the decoding instance has inter-op = 3, intra-op = 4. Under this placement, \sys can effectively balance the load between the two instances on ShareGPT, meeting latency requirements at the lowest cost. This non-trivial placement strategy is challenging to manually find, proving the effectiveness of the algorithm. In the case of vLLM, collocating prefill and decoding greatly slows down the decoding phase, thereby significantly increasing TPOT. Due to the stringent TPOT requirements of chatbot applications, although vLLM meets the TTFT SLO for most requests, the overall SLO attainment is dragged down by a large number of requests that violate the TPOT SLO. Compared to DeepSpeed-MII, \sys can sustain $1.6\times$--$7.4\times$ higher request rate. DeepSpeed-MII shows better performance on larger models because the prefill job is larger and chunked-prefill mitigates the interference to some extent. However, due to the reasons discussed in \S\ref{sec:problem-opportunity}, chunked prefill is slower than full prefill, so it struggles to meet the TTFT SLO as a sacrifice for better TPOT.

The second row of Figure~\ref{fig:chatbot} indicates the robustness to the changing latency requirements of the two systems. We fix the rate and then linearly scale the two latency requirements in Table~\ref{tab:app_config} simultaneously using a parameter called \textit{SLO Scale}. As SLO Scale decreases, the latency requirement is more stringent. We aim to observe the most stringent SLO Scale that the system can withstand while still achieving the attainment target. Figure~\ref{fig:chatbot} shows that \sys can achieve $1.8\times$--$3.2\times$ more stringent SLO than vLLM and $1.7\times$--$1.8\times$ more stringent SLO than DeepSpeed-MII, thus providing more engaging service quality to the users.

\parabf{Code completion.} Figure~\ref{fig:other_apps}\textcolor{green!80!black}{(a)} shows the performance of \sys on the code completion task when serving OPT-66B. \sys can sustain $5.7\times$ higher request rate and $1.4\times$ more stringent SLO than vLLM. Compared to DeepSpeed-MII, \sys can sustain $1.6\times$ higher request rate and $1.4\times$ more stringent SLO. As a real-time coding assistant, the code completion task demands lower TTFT than chatbot, this leads to both systems ultimately being constrained by the TTFT requirement. However, in comparison, by eliminating the interference of the decoding jobs and automatically increasing intra-operation parallelism in prefill instances through the searching algorithm, \sys reduces the average latency of the prefill jobs, thereby meeting the TTFT requirements of more requests. 

\parabf{Summarization.} Figure~\ref{fig:other_apps}\textcolor{green!80!black}{(b)} shows the performance of \sys on the summarization task when serving OPT-66B. \sys achieves $4.3\times$ higher request rate and $12.6\times$ more stringent SLO than vLLM. Compared to DeepSpeed-MII, \sys achieves $1.8\times$ higher request rate and $2.6\times$ more stringent SLO. The requests sampled from LongBench dataset have long input lengths, which brings significant pressure to the prefill computation. However, due to the loose requirement of TTFT for the summarization task, the TPOT service quality becomes particularly important. Since vLLM collocates prefill and decoding phases, it experiences a greater slowdown in the decoding phase with long prefill jobs and fails to meet the TPOT requirement.

The results above are all under the 90\% SLO attainment target. We observe that \sys can have better performance under a more stringent attainment target (say, 99\%) and present the results in Appendix~\ref{appendix:end_to_end_p99}.

\begin{figure}[!t]
    \centering
    \includegraphics[width=\columnwidth]{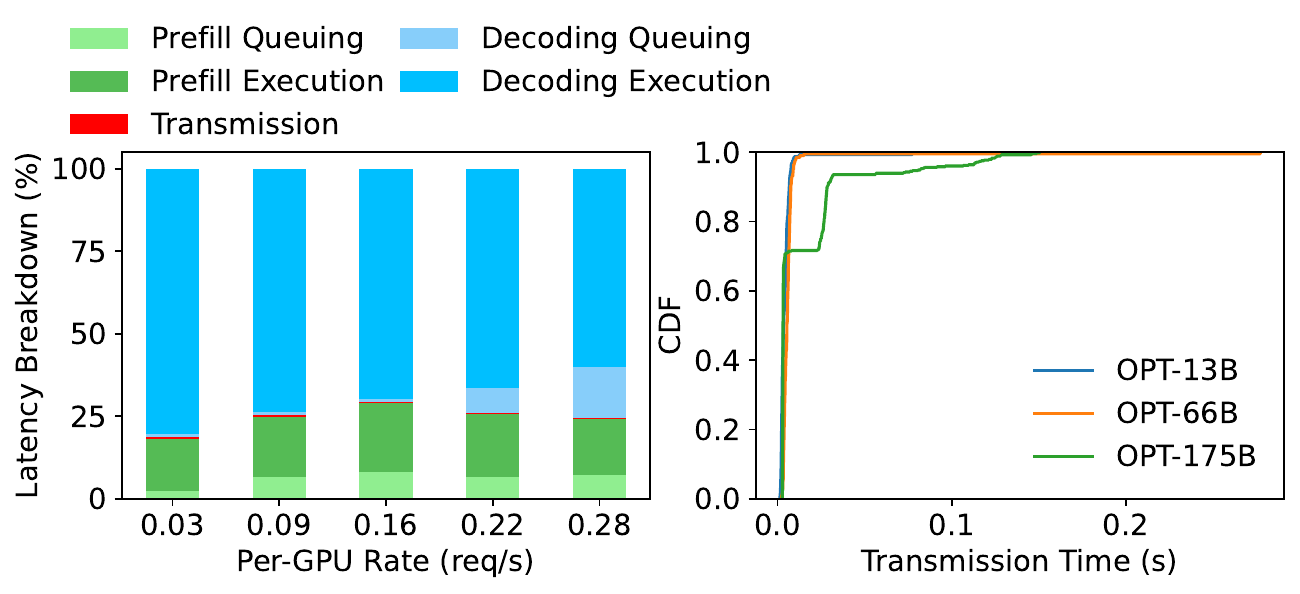}
    \caption{\textit{Left:} Latency breakdown when serving OPT-175B on ShareGPT dataset with \sys. \textit{Right:} The CDF function of KV Cache transmission time for three OPT models.}
    \vspace{-4mm}
    \label{fig:overhead}
\end{figure}

\subsection{Latency Breakdown}
\label{exp:breakdown}

To understand \sys's performance in detail, we make a latency breakdown of the requests in \sys. We divide the processing lifecycle of a request in \sys into five stages: \textit{prefill queuing}, \textit{prefill execution}, \textit{transmission}, \textit{decoding queuing}, and \textit{decoding execution}. The total time consumed by all requests in each stage is then summed up to determine their respective proportions in the system's total execution time. 

Figure~\ref{fig:overhead}\textcolor{green!80!black}{(a)} shows the latency breakdown for the OPT-175B models on the ShareGPT dataset. We chose OPT-175B because the KV Cache transmission is more demanding for larger models. In fact, even for OPT-175B, the KV Cache transmission only accounts for less than 0.1\% of the total latency. Even by examining the CDF of the absolute transmission time shown in Figure~\ref{fig:overhead}\textcolor{green!80!black}{(b)}, we observe that over 95\% of requests experience a delay of less than 30ms, despite our testbed having only limited cross-node bandwidth. This is due to the algorithm described in \S\ref{subsec:practical_placement}, where we require the prefill and decoding instance to maintain the same stage on one machine, enabling the use of intra-node NVLINK bandwidth for transmission, thus significantly reducing transmission delay. 

\begin{table}
\footnotesize
\centering
\begin{tabular}{c|cc|cc}
\toprule
\multirow{2}{*}{\begin{tabular}[c]{@{}c@{}}Rate\\ (req/s)\end{tabular}} & \multicolumn{2}{c|}{vLLM} & \multicolumn{2}{c}{\sys-Low} \\  \cline{2-3} \cline{4-5}
      & Real System & Simulator & Real System & Simulator \\ \midrule
1.0    & 97.0\%      & 96.8\%    & 100.0\%     & 100.0\%    \\
1.5    & 65.5\%      & 65.1\%    & 100.0\%     & 100.0\%    \\
2.0    & 52.8\%      & 51.0\%    & 99.3\%      & 99.3\%    \\
2.5    & 44.9\%      & 46.1\%    & 87.3\%      & 88.3\%    \\
3.0    & 36.7\%      & 38.3\%    & 83.0\%      & 84.1\%    \\
3.5    & 27.8\%      & 28.0\%    & 77.3\%      & 77.0\%    \\
4.0    & 23.6\%      & 24.1\%    & 70.0\%      & 68.9\%    \\
\bottomrule
\end{tabular}
\vspace{-2mm}
\caption{Comparison of the SLO attainment reported by the simulator and the real system under different rates.}
\vspace{-4mm}
\label{table:simulator_fidelity}
\end{table}

\begin{figure}[!t]
    \centering
    \includegraphics[width=\columnwidth]{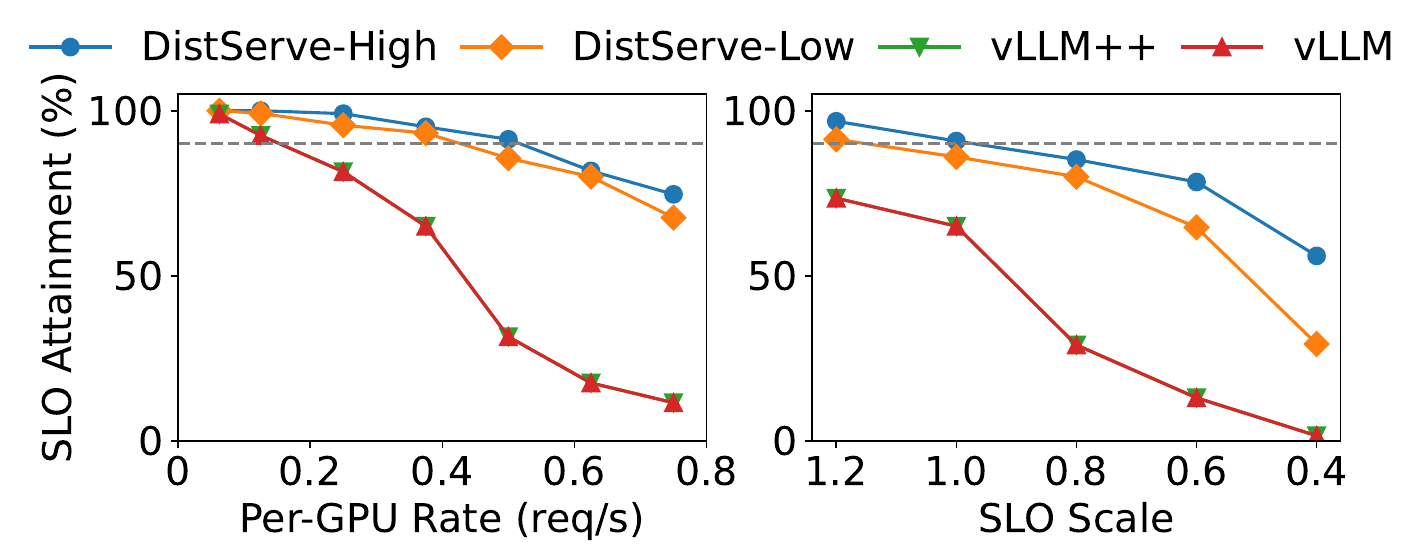}
    \vspace{-6mm}
    \caption{Ablation experiments.}
    \vspace{-4mm}
    \label{fig:ablation}
\end{figure}

\subsection{Ablation Studies}
\label{exp:ablation}
We study the effectiveness of the two key innovations in \sys: disaggregation and the placement searching algorithm. In \S\ref{exp:end_to_end}, we choose the default parallelism setting for vLLM following its original paper~\cite{vllm}. So we implement "vLLM++" which enumerates different parallelism strategies and chooses the best. For \sys, We also compare the placement found by Alg.~\ref{alg:low-affinity-placement} (\sys-Low) with the one found by Alg.~\ref{alg:optimal-placement} (\sys-High) which has fewer searching constraints and assumes high cross-node bandwidth. Since vLLM does not support inter-op parallelism and our physical testbed does not have high cross-node bandwidth, we use simulation for this experiment.

\parabf{Simulator accuracy.} Noticing that DNN model execution~\cite{gujarati2020serving} has high predictability, even under parallel settings~\cite{li2023alpaserve, alpa}.
We study the accuracy of the simulator in Tab.~\ref{table:simulator_fidelity}. For "vLLM" and "\sys-Low", we compare the SLO attainment reported by the simulator and by real runs on our testbed under different rates. The error is less than 2\% in all cases, verifying the accuracy of our simulator.

\parabf{Results.} Figure~\ref{fig:ablation} shows the performance of the four systems when serving OPT-66B on the ShareGPT dataset. "vLLM++" has the same performance as "vLLM" because we find the default parallelism setting (intra-op=4) has the best per-GPU goodput. This further demonstrates the importance of disaggregation. The interference between the prefill and decoding phases significantly reduces the potential performance improvement through adjusting parallelism. In contrast, "DistLLM-High" can achieve further improvements over "DistLLM-Low" because it is not constrained by the deployment constraint that the prefill and decoding instance on one node should share the same model stage. Through disaggregation, we can use tailored parallelism strategies for prefill and decoding instances and optimize their targets without the coupling effects.

\begin{figure}[!t]
    \centering
    \includegraphics[width=0.8\columnwidth]{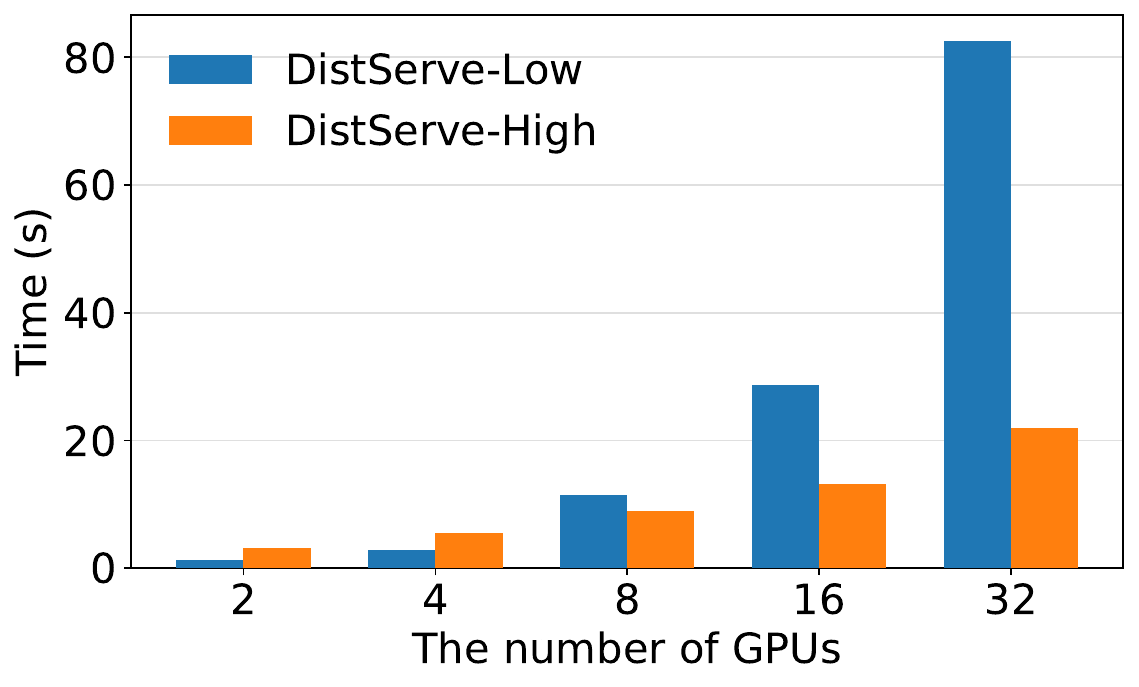}
    \vspace{-3mm}
    \caption{Algorithm Running Time}
    \label{fig:runtime}
    \vspace{-5mm}
\end{figure}

\subsection{Algorithm Running Time}
\label{exp:run_time}
Figure~\ref{fig:runtime} shows the running time for Alg.~\ref{alg:optimal-placement} (\sys-Low) and Alg.~\ref{alg:low-affinity-placement} (\sys-High) on an AWS m5d.metal instance with 96 cores as the number of GPUs ($N \times M$) provided to a single instance increases. According to the results, \sys scales well with the number of GPUs and is independent of the model size. This is because the simulator only simulates discrete events and the running time is the same no matter how big the model is. On the other hand, both algorithms are highly parallelizable, as the searches for different parallelism strategies are independent of each other, allowing the execution time of the algorithms to accelerate almost linearly with more CPU cores. 

As the number of GPUs increases, the execution time of "Dist-Low" becomes higher than that of "Dist-High". This is because the search for parallelism strategies for prefill and decoding instances in "Dist-High" is independent and can be parallelized. But for "Dist-Low", due to additional restrictions on deployment, we need to enumerate all the possible intra-node parallelism combinations for prefill and decoding instances. Even so, the execution time of the algorithm is in minutes, and since it only needs to be executed once before each redeployment, this overhead is acceptable.

\section{Discussion}
\label{sec:discussion}
In this paper, we focus on the goodput-optimized setting and propose \sys under the large-scale LLM serving scenario. As LLMs are widely used and deployed across various service scenarios with different optimization targets and resource limits, it becomes almost impossible to find a one-size-fits-all solution that effectively addresses all aspects of LLM serving. In this section, we discuss the pros and cons of \sys and potentially better solutions in other scenarios.

\parabf{Throughput-optimized scenarios.} In offline applications that are not latency-sensitive, users typically have lower requirements for response time~\cite{sheng2023flexgen}. This allows serving systems to shift focus towards maximizing overall throughput instead of goodput and the effectiveness of DistServe may be compromised. In this case, techniques such as chunked-prefill with piggyback~\cite{agrawal2023sarathi,deepspeed_mii} may be preferred since it can fill each batch to the compute-bound threshold, thereby maintaining higher GPU utilization in every iteration.

\parabf{Resource-constrained scenarios.} Small-scale enterprises and individual researchers often lack the resources to deploy LLMs on large-scale clusters~\cite{sheng2023flexgen, song2023powerinfer}. In resource-constrained scenarios, such as environments with only a few or even a single GPU, the design space for DistServe is significantly limited. It struggles or even fails to adjust the parallel strategies and resource allocation to effectively enhance serving performance. In this case, simpler architectural choices like non-disaggregated systems~\cite{vllm, deepspeed_mii} may reduce deployment complexity and optimize operational efficiency.

\parabf{Long-context scenarios.} Nowadays, more and more GPT models support extremely long contexts, such as Claude-3~\cite{claude3}, Gemini-1.5~\cite{gemini}, and Large World Model (LWM)~\cite{liu2023world}, which all have a 1M context window. In such scenarios, the transmission overhead will increase as the size of the KV cache grows linearly with the prompt length. However, the prefill computation grows quadratically, so the relative duration of transmission and prefill job decreases. Meanwhile, a longer context further exacerbates the disparity in computational demands between prefill and decoding jobs, leading to increased interference between them. Therefore, the disaggregation approach proposed in \sys remains promising in long-context serving.
\section{Related Work}
\label{sec:related}

\parabf{Inference serving.} There has been plenty of work on inference serving recently. They range from general-purpose production-grade systems like TorchServe~\cite{torchserve} and NVIDIA Triton~\cite{nvidiatriton} to systems optimized specifically for Transformer-based LLMs~\cite{li2023alpaserve, yu2022orca, agrawal2023sarathi, wu2023fast, fang2021turbotransformers,fastertransformer,zhou2022pets,su2023hotgpt}. 
Among them, Orca~\cite{yu2022orca} introduces continuous batching to increase throughput. vLLM~\cite{vllm} proposes paged-attention for fine-grained KV cache management. SARATHI~\cite{agrawal2023sarathi} suggests a chunked-prefill approach, splitting a prefill request into chunks and piggybacking decoding requests to improve hardware utilization. FastServe~\cite{wu2023fast} implements iteration-level preemptive scheduling to mitigate the queuing delay caused by long jobs. However, they all employ a colocation approach for prefill and decoding processing, thus leading to severe interference. There are also concurrent works such as Splitwise~\cite{patel2023splitwise}, TetriInfer~\cite{hu2024inference} and DéjàVu~\cite{strati2024dejavu} which adopt similar disaggregation idea to optimize LLM inference, further confirming the effectiveness of this method. Differently, \sys emphasizes the goodput optimization scenario more and takes a closer look at the aspect of network bandwidth.

\parabf{Goodput-optimized systems.} Optimizing goodput is a hot topic in DL applications. Pollux~\cite{pollux} improves scheduling performance in DL clusters by dynamically adjusting resources for jobs to increase cluster-wide goodput. Sia~\cite{jayaram2023sia} introduces a heterogeneous-aware scheduling approach that can efficiently match cluster resources to elastic resource-adaptive jobs. Clockwork~\cite{clockwork} and Shepherd~\cite{zhangshepherd} provide latency-aware scheduling and preemption to improve the serving goodput, but they only target traditional small models. AlpaServe~\cite{li2023alpaserve} focuses on LLMs, employing model parallelism to statistically multiplex the GPU execution thus improving the resource utilization. However, it only targets the non-autoregressive generation. \sys is the first work to optimize the goodput for autoregressive LLM inference.

\parabf{Resource disaggregation.} Resource disaggregated systems~\cite{shan2018legoos,mira,CXL} decouple the hardware resources from the traditional monolithic server infrastructure and separate them into resource pools to manage independently. It allows for more flexible, efficient, and scalable deployment and increases resource utilization. Many applications benefit from a truly disaggregated data center with high-speed network bandwidth and heterogenous hardware support ~\cite{makeitreal,audibert2022case, distmind}. \sys shares the concept by disaggregating its system components, allowing for independent resource scaling and management.

\parabf{Model parallelism for training.} \sys is orthogonal to the large body of work on model parallelism in training~\cite{alpa,shoeybi2020megatronlm,huang2019gpipe,rajbhandari2020zero,pipedream}. As described in \S\ref{subsec:practical_problems}, inference-serving workloads have unique characteristics not found in training settings. Where these systems do intersect with \sys, is in their methods for implementing model parallelism along various dimensions. \sys can integrate new parallelism optimizations into its placement searching algorithm.

\section{Conclusion}
\label{sec:conclusion}
We present \sys, a new LLM serving architecture that disaggregates the prefill and decoding computation. \sys maximizes the per-gpu goodput -- the maximum request rate that can be served adhering to the SLO attainment goal for each GPU provisioned, hence resulting in up to $7.4\times$ lower cost per LLM query with guaranteed satisfaction of SLOs. 
Our findings affirm that as latency becomes an increasingly important metric for LLM services, prefill and decoding disaggregation is a vital strategy in promising improved performance and service quality guarantees.

\parabf{Acknowledgments.} We sincerely thank our shepherd and
the anonymous reviewers for their valuable feedback. This work was
supported by the National Natural Science Foundation of China under the grant numbers
62172008, 62325201, and the National Natural Science Fund for the Excellent Young Scientists Fund
Program (Overseas). Junda Chen is supported by UCSD fellowship and Hao Zhang is supported by UCSD faculty startup fund. Xin Jin is
the corresponding author. Yinmin Zhong, Xuanzhe Liu, and Xin Jin are
also with the Key Laboratory of High Confidence Software Technologies (Peking
University), Ministry of Education.

\balance
\bibliographystyle{plain}
\bibliography{paper.bib}

\newpage
\appendix

\section{Latency Model for LLM Inference}
\label{appendix:latency_model}

To accurately simulate the goodput of different placement strategies, we use an analytical model to predict the execution time of the prefill and decoding phases in LLM inference. 

In modern LLM serving systems~\cite{fastertransformer, vllm, wu2023fast}, memory-bound operations like Softmax and LayerNorm are usually fused with matrix multiplication kernels for efficiency. Thus the GEMMs dominate the overall latency and our analysis primarily focuses on them. 

\subsection{Symbol Definition}

Here are symbols related to the architecture of the model:

\begin{itemize}
    \item $h$: hidden size
    \item $n$: number of heads
    \item $s$: head size ($h = n \cdot s$)
    \item $m$: FFN intermediate size
\end{itemize}

Note: If tensor parallelism is used, $h$, $n$, and $m$ should be divided by the tensor parallelism size.

Below are symbols that characterize the batch to be executed:

\begin{itemize}
    \item $B$: batch size
    \item $l_0, l_1, \dots, l_{B-1}$: input length of each request within the batch
    \item $t$: number of tokens in the batch, ($t = \sum_{i=0}^{B-1} l_i$)
    \item $t_2$: squared sum of the input lengths ($t_2 = \sum_{i=0}^{B-1} l_i^2$)
    \item $b$: block size in the attention kernel. This parameter is used in FlashAttention~\cite{dao2022flashattention}, a common kernel optimization technique adopted by current LLM serving systems.
\end{itemize}

\subsection{Prefill Phase Latency Modeling}

Since the attention operation uses specially optimized kernels, we first discuss the other four matrix multiplications in the prefill phase:

\begin{table}[h]
\centering
\begin{tabular}{|c|c|c|}
    \hline
    GEMM Name & Shape of $M$ & Shape of $N$ \\ \hline
    QKV Linear & $(t, h)$ & $(h, 3h)$ \\ \hline
    Attn Output & $(t, h)$ & $(h, h)$ \\ \hline
    FFN Input & $(t, h)$ & $(h, m)$ \\ \hline
    FFN Output & $(t, m)$ & $(m, h)$ \\ \hline
\end{tabular}
\end{table}

The arithmetic intensity (AI) of these operations is $O(t)$. On NVIDIA A100-80GB GPU, it is compute-bound when AI is over 156. Since $t$ usually can reach several hundred in real cases, all of these operations are compute-bound. Therefore, we can model the latency of these operations according to the total FLOPs:

$$T_1 = C_1 \cdot (4th^2 + 2thm)$$

Next, we discuss the prefill attention operation with FlashAttention\cite{dao2022flashattention} optimization. Since the attention only operates among the tokens in the same request, current implementations launch attention kernels for each request in the same batch. For one attention head and a request with $l$ tokens, the attention kernel needs to perform a total of $2sl + 3sl\cdot(l/b) \approx 3sl\cdot(l/b)$ memory reads and writes, alongside $2sl^2 + sl(l/b) \approx 2sl^2$ FLOPs. So the AI is $2b/3 = 10.677$ (when $b=16$) or $21.333$ (when $b=32$), indicating that it is a memory-bound operation on A100 GPU. Therefore, the whole attention layer latency (including all requests and all heads) can be modeled as:

$$T_2 = C_2 \cdot n \cdot \sum_{i=0}^{B-1} \frac{3sl_i^2}{b} = C_2 \cdot \frac{3nst_2}{b} = C_2 \cdot \frac{3ht_2}{b}$$

Overall, the latency of the prefill phase can be modeled as:

$$T_{Prefill} = C_1 \cdot (4th^2 + 2thm) + C_2 \cdot \frac{3ht_2}{b} + C_3$$

We use $C_3$ to quantify other overheads like Python Runtime, system noise, and so on. Then we use profiling and interpolation to figure out the values of $C_1$, $C_2$, and $C_3$.

\subsection{Decoding Phase Latency Modeling}

Similarly, we first focus on the following GEMMs in the decoding phase:

\begin{table}[h]
\centering
\begin{tabular}{|c|c|c|}
    \hline
    GEMM Name & Shape of $M$ & Shape of $N$ \\ \hline
    QKV Linear & $(B, h)$ & $(h, 3h)$ \\ \hline
    Attn Output & $(B, h)$ & $(h, h)$ \\ \hline
    FFN Input & $(B, h)$ & $(h, m)$ \\ \hline
    FFN Output & $(B, m)$ & $(m, h)$ \\ \hline
\end{tabular}
\end{table}

The AI of these operations is $O(B)$. $B$ is limited by the GPU memory size and stringent latency requirements, so in existing serving scenarios, these operations are memory-bound. The total memory reads and writes is $8Bh + 4h^2 + 2hm + 2Bm$, and since $h$ and $m$ are usually significantly larger than $B$, we can model the latency as:

$$T_3 = C_4 \cdot (4h^2 + 2hm)$$

As for the decoding attention operation, for one attention head and a request with $l$ generated tokens, it needs to perform $3sl$ memory reads and writes, alongside $2sl$ FLOPs. It is memory-bound, so we can model the latency of decoding attention as:

$$T_4 = C_5 \cdot n \cdot 3s \sum_{i=0}^{B-1} l_i = C_5 \cdot 3ht$$

Summing up, the latency of the decoding phase is:

$$ T_{Decoding} = C_4 \cdot (4h^2 + 2hm) + C_5 \cdot 3ht$$

Here we do not introduce the overhead term (like $C_3$ in the profiling stage) because $4h^2 + 2hm$ is already a constant, and the overhead can be put into $C_4$. Similarly, we use profiling and interpolation to figure out the values of $C_4$ and $C_5$.

\begin{figure*}[!t]
    \centering
    \includegraphics[width=0.95\linewidth]{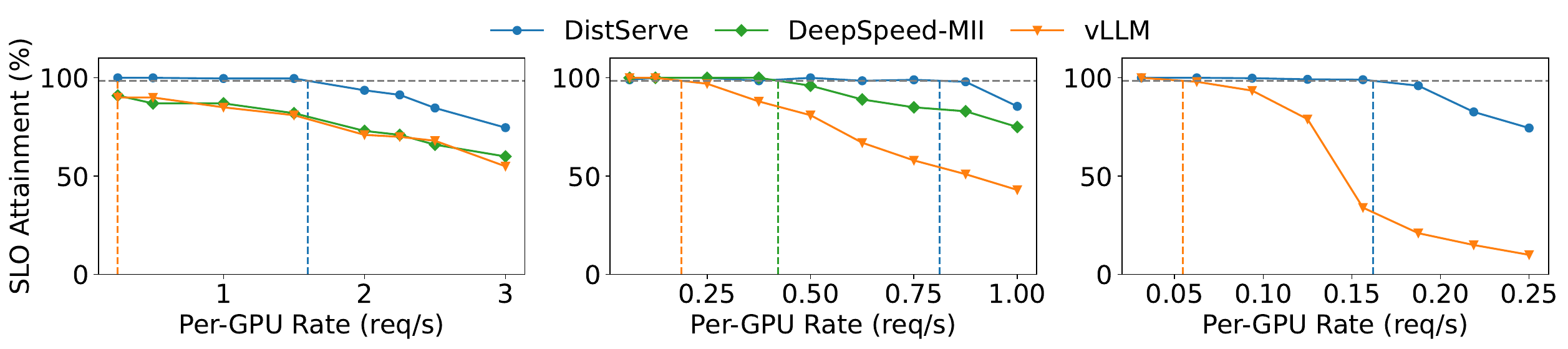}
    \includegraphics[width=0.95\linewidth]{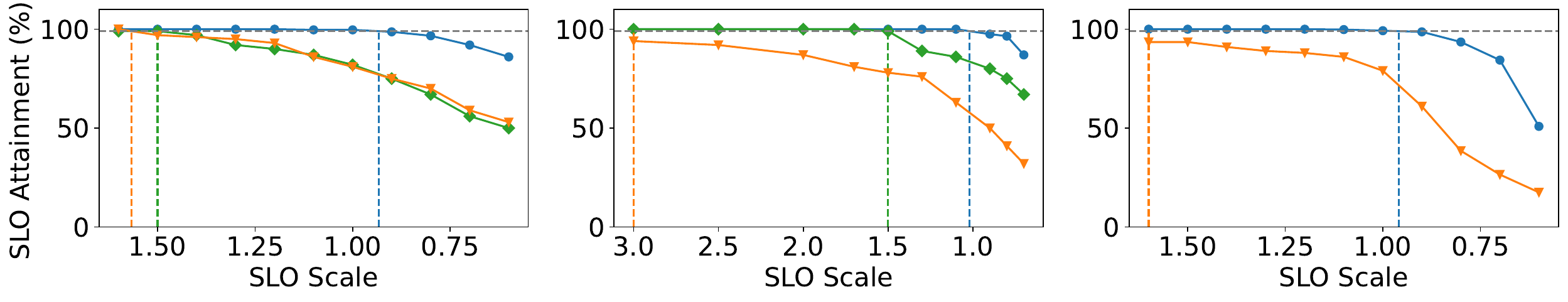}
    \hspace{10mm} (a) OPT-13B \hspace{40mm} (b) OPT-66B \hspace{40mm} (C) OPT-175B
    \caption{Chatbot application with OPT models on the ShareGPT dataset.}
    \label{fig:chatbot_p99}
\end{figure*}

\begin{figure*}[!t]
    \centering
    \includegraphics[width=0.95\linewidth]{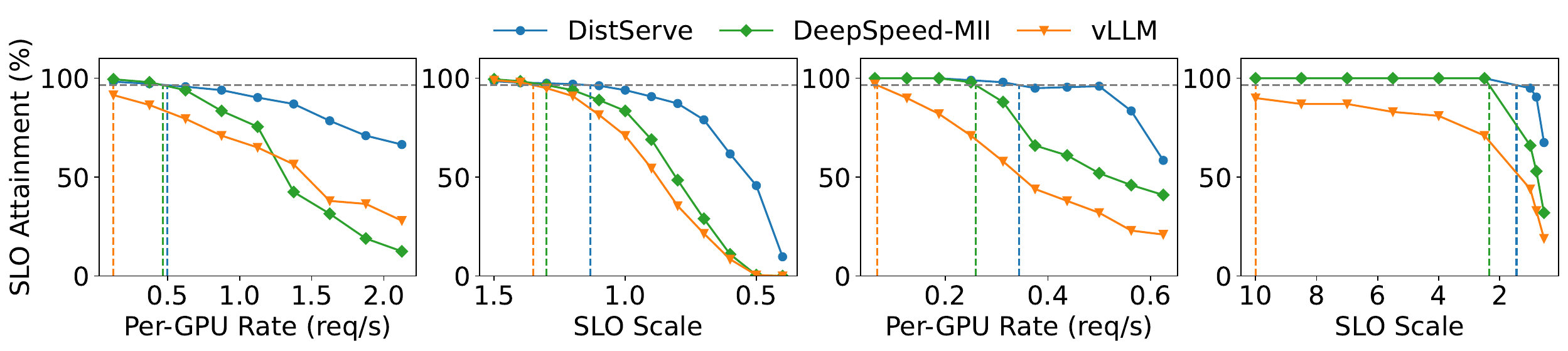}
    \hspace{10mm} (a) Code Completion \hspace{50mm} (b) Summarization
    \caption{Code completion and summarization tasks with OPT-66B on HumanEval and LongBench datasets, respectively.}
    \label{fig:other_apps_p99}
\end{figure*}

\section{\sys Placements in End-to-end Experiments}
\label{appendix:placement_strategies}

Table~\ref{tab:parallel_config} shows the tensor parallelism (TP) and pipeline parallelism (PP) configurations for prefill and decoding instances chosen by \sys in the end-to-end experiments \S\ref{exp:end_to_end}.

\begin{table}[h]
\begin{tabular}{|c|c|c|c|c|c|}
    \hline
    \multirow{2}{*}{Model} & \multirow{2}{*}{Dataset} & \multicolumn{2}{c|}{Prefill} & \multicolumn{2}{c|}{Decoding} \\ \cline{3-6}
    & & TP & PP & TP & PP \\ \hline
    OPT-13B & ShareGPT & 2 & 1 & 1 & 1 \\ \hline
    OPT-66B & ShareGPT & 4 & 1 & 2 & 2 \\ \hline
    OPT-66B & LongBench & 4 & 1 & 2 & 2 \\ \hline
    OPT-66B & HumanEval & 4 & 1 & 2 & 2 \\ \hline
    OPT-175B & ShareGPT & 3 & 3 & 4 & 3 \\ \hline
\end{tabular}
\caption{The parallelism strategies chosen by \sys in the end-to-end experiments.}
\label{tab:parallel_config}
\end{table}

\section{End-to-end Results under 99\% SLO attainment}
\label{appendix:end_to_end_p99}
Figure~\ref{fig:chatbot_p99} and Figure~\ref{fig:other_apps_p99} show the end-to-end performance between \sys and baselines with the same setup in \S\ref{exp:end_to_end} except that the SLO attainment goal is changed to 99\%. We can see that under a more stringent SLO attainment goal, compared to vLLM, \sys can still sustain $3\times$--$8\times$ higher rate and $1.24\times$--$6.67\times$ more stringent SLO. When compared to DeepSpeed-MII, \sys can achieve $1.32\times$--$8\times$ higher rate and $1.20\times$--$1.58\times$ more stringent SLO.

\end{document}